\documentclass[preprint]{aastex631}

\usepackage{lineno}

\shorttitle{Annotated Coadds}
\graphicspath{{./}{figures/}}
\begin{document}
%\linenumbers
\title{Annotated Coadds: Concise Metrics for Characterizing Survey Cadence and
for Discovering Variable and Transient Sources}

%% LaTeX will automatically break titles if they run longer than
%% one line. However, you may use \\ to force a line break if
%% you desire. In v6.3 you can include a footnote in the title.

%% The new \altaffiliation can be used to indicate some secondary information
%% such as fellowships. This command produces a non-numeric footnote that is
%% set away from the numeric \affiliation footnotes.  NOTE that if an
%% \altaffiliation command is used it must come BEFORE the \affiliation call,
%% right after the \author command, in order to place the footnotes in
%% the proper location.
%%
%% Use \email to set provide email addresses. Each \email will appear on its
%% own line so you can put multiple email address in one \email call. A new
%% \correspondingauthor command is available in V6.3 to identify the
%% corresponding author of the manuscript. It is the author's responsibility
%% to make sure this name is also in the author list.
%%
%% While authors can be grouped inside the same \author and \affiliation
%% commands it is better to have a single author for each. This allows for
%% one to exploit all the new benefits and should make book-keeping easier.
%%
%% If done correctly the peer review system will be able to
%% automatically put the author and affiliation information from the manuscript
%% and save the corresponding author the trouble of entering it by hand.

\correspondingauthor{David L. Shupe, Ranga Ram Chary}
\email{shupe@ipac.caltech.edu, rchary@ipac.caltech.edu}

\author[0000-0003-4401-0430]{David L. Shupe}
\affiliation{IPAC, California Institute of Technology, 1200 E. California Blvd, Pasadena, CA 91125, USA}
             
\author[0000-0002-8532-9395]{Frank J. Masci}
\affiliation{IPAC, California Institute of Technology, 1200 E. California Blvd, Pasadena, CA 91125, USA}

\author[0000-0001-7583-0621]{Ranga Ram Chary}
\affiliation{IPAC, California Institute of Technology, 1200 E. California Blvd, Pasadena, CA 91125, USA}

\author[0000-0003-3367-3415]{George Helou}
\affiliation{IPAC, California Institute of Technology, 1200 E. California Blvd, Pasadena, CA 91125, USA}

\author[0000-0002-9382-9832]{Andreas Faisst}
\affiliation{IPAC, California Institute of Technology, 1200 E. California Blvd, Pasadena, CA 91125, USA}
             
\author[0000-0002-0077-2305]{Roc M. Cutri}
\affiliation{IPAC, California Institute of Technology, 1200 E. California Blvd, Pasadena, CA 91125, USA}

\author{Tim Brooke}
\affiliation{IPAC, California Institute of Technology, 1200 E. California Blvd, Pasadena, CA 91125, USA}

\author[0000-0001-7291-0087]{Jason A Surace}
\affiliation{IPAC, California Institute of Technology, 1200 E. California Blvd, Pasadena, CA 91125, USA}  

\author[0000-0003-0107-7803]{Ken A Marsh}
\affiliation{IPAC, California Institute of Technology, 1200 E. California Blvd, Pasadena, CA 91125, USA}

%% Mark off the abstract in the ``abstract'' environment. 
\begin{abstract}

In order to study transient phenomena in the Universe, existing and forthcoming imaging surveys are covering wide areas of sky repeatedly
over time, with a range of cadences, point spread functions, and depths. We describe here a framework that allows
an efficient search for different types of time-varying astrophysical phenomena
in current and future, large data repositories. 
We first present a methodology to generate and store key survey parameters that enable researchers to determine
if a survey, or a combination of surveys, allows specific time-variable astrophysical phenomena to be
discovered. To facilitate further exploration of sources in regions of interest, we then generate a few sample
metrics that capture the essential brightness characteristics of a sky pixel at a specific wavelength. Together, we refer to these as ``annotated coadds".
The techniques presented here for WISE/NEOWISE-R data are sensitive to 10\% brightness variations at around 12th Vega magnitude at 4.5 microns wavelength. Application of the technique to ZTF data
also enabled the detection of 0.5 mag variability at 20 AB mag in the $r-$band. We demonstrate the capabilities of these metrics for different classes of sources: high proper-motion stars, periodic variable stars, and supernovae, and find that each metric has its advantages depending on the nature of variability. We also present a data structure which will ease the search for temporally varying phenomena in future surveys.
\end{abstract}

%% Keywords should appear after the \end{abstract} command. 
%% See the online documentation for the full list of available subject
%% keywords and the rules for their use.

%% From the front matter, we move on to the body of the paper.
%% Sections are demarcated by \section and \subsection, respectively.
%% Observe the use of the LaTeX \label
%% command after the \subsection to give a symbolic KEY to the
%% subsection for cross-referencing in a \ref command.
%% You can use LaTeX's \ref and \label commands to keep track of
%% cross-references to sections, equations, tables, and figures.
%% That way, if you change the order of any elements, LaTeX will
%% automatically renumber them.
%%
%% We recommend that authors also use the natbib \citep
%% and \citet commands to identify citations.  The citations are
%% tied to the reference list via symbolic KEYs. The KEY corresponds
%% to the KEY in the \bibitem in the reference list below. 

\section{Introduction} \label{sec:intro}
Projects such as the NEOWISE Reactivation mission (NEOWISE-R), the Palomar Transient Factory (PTF), the Zwicky Transient Factory (ZTF) and in the future, the Vera C. Rubin Observatory, the Euclid observatory,
the Nancy Grace Roman Telescope, and the NEO Surveyor mission will be covering wide areas of
sky repeatedly, with various cadences. They will produce hundreds of petabytes
of imaging data in which to search for temporally varying phenomena, a challenge for
traditional astronomical analysis techniques. Time-varying astrophysical phenomena
can be broadly classified into three categories: 

\begin{enumerate}
\item Flux-variable sources at a fixed position on the sky, such as active
   galactic nuclei (AGN), pulsating variable stars, eclipsing binary stars,
   and a myriad of transient phenomena: flare stars, supernovae (SNe), microlensing events,
   and electromagnetic signatures from mergers between black holes and/or
   neutron stars; 

\item Sources constant in flux but which are varying in position on the sky, such as stars with proper motion (and/or parallax) and certain classes of asteroids moving at relatively slow angular rates on the sky, for example, Trans-Neptunian Objects (TNOs) and Main Belt Comets (MBCs) or comets in the outer main belt and beyond;

\item Sources which are both varying in flux and position such as asteroids and comets.  

\end{enumerate}

The science drivers to identify and monitor these classes of sources are many and
varied. Detection and classification of a homogeneous sample of Type Ia SNe can measure cosmological parameters and the nature of
the dark energy which makes up $\sim$70\% of the Universe \citep[e.g][]{Perlmutter1997}. The duty cycle of AGN \citep[e.g][]{Prakash} as measured
by brightness variations can reveal the growth of supermassive black holes, and its impact on the regulation of stellar mass in the host galaxy. Identification of asteroids and comets in the outer Solar System, for example TNOs (which move at slow angular rates) can help
characterize their distribution and composition, and constrain evolutionary models of the Solar System. Tracking the brightness variation of comets as a function of
heliocentric distance can reveal their composition, as well as the fraction of the zodiacal dust cloud which is produced by cometary out-gassing \citep[e.g.][]{Jewitt}. 

Transient detection has typically required creating a deep reference image template,
astrometrically and photometrically matching new epochal images to this template, and
subtracting it from the single epoch images. For ground-based imaging in
particular, one also needs to match point spread functions (PSFs) due to
atmospheric seeing variations \citep[e.g.,][]{zogy}. Space-based image
data may also have systematically varying PSFs relative to a fixed reference
image, for example due to asymmetries in the PSF coupled with rotation of the instrument
field-of-view (FOV) as well as non-isoplanicity across the FOV. 
The process typically results in a large number of spurious sources not only from
instrumental artifacts and residuals from imperfect subtraction, but also
cosmic rays and satellite trails. The former affect all exposures while the latter probably
affect 1\% of exposures even in space-based data sets. This results in a very large
volume of data to search for temporally variable phenomena, which can be prohibitive,
especially for fast searches in archives which are hosted at different data centers.

This paper outlines steps towards streamlining the identification of time-domain astronomical phenomena.
Before such phenomena can be identified, it is necessary to encapsulate the characteristics of an observational survey in such a way that it can be searched for time-domain phenomena. For example, if a particular survey does not even have repeated observations of the same patch of sky, or does not overlap in time with a particular unusual astrophysical phenomenon, clearly it is inappropriate for identifying variability.
These survey characteristics can be stored in maps which have relatively coarse spatial resolution. Once this information is generated, we can generate quantitative metrics that capture the essential characteristics of the historical
behavior of a sky pixel at a given wavelength. We call these two products together ``annotated coadds" since they facilitate
searches for variable phenomena.
Relevant metrics must cover two broad variability timescales: ``fast''
transients and slow or repeating trends (either periodic or aperiodic). 

Identification of fast transients in new data needs to be achieved with a short turnaround time.
A typical use case here may be identifying candidate new transients in the many square degrees \citep{LVC2020} LIGO-VIRGO gravitational wave localization region - by comparing new imaging data with variability metrics for the same patch of sky, other known variable sources can be filtered out. Another use case could be the flaring of a blazar which could potentially be the source of ultra high-energy neutrinos \citep{Buson, Icecube}. Identifying all variable AGN/blazars coincident with neutrino events could potentially be accomplished by searching the optical/infrared annotated coadds for pixels with high variability and cross-matching them with X-ray catalogs. 
Finally, the variability metrics can be used to find missed
transients to facilitate completeness studies.

Studying slow trends relies on analysis
of the time history, covering potentially hundreds of visits, to characterize
temporal behavior, usually with specialized techniques depending on the phenomena sought. While assessing whether a source is variable or not can be abbreviated into a single parameter,  generating precise light curves will require forced photometry on the individual epochal images - this information can be stored for every variable source dramatically reducing the volume of data that needs to be searched. In either case, the historical behavior
of a given sky location is the first question asked when a transient is
detected, to help constrain its nature. 

Here, we outline an investigation of several procedures to generate annotated coadds
with the goal of designing and developing general purpose software to be used for the archiving and searching of data from any
astronomical survey sensitive to time-domain phenomena. We have studied the
applicability of several metrics to various examples drawn from WISE \citep{Wright2010} and NEOWISE-R \citep{Mainzer_2014}
data with over a hundred typical visits at every sky point. We have also investigated the application of some of these metrics to r-band imaging data from ZTF \citep{Bellm2019,Masci2019}.  

\section{Sky Tessellation}

The first step is to encode characteristics of an observational survey. Even a single mission like {\it Spitzer} or {\it WISE} can have both time-domain components and wide-area, single-visit components. Even identifying which areas of sky are suitable for searches for time-domain phenomena requires selecting a pixel basis (geometry) for data storage. Many different choices exist. Wide-area surveys, such as with Subaru/HSC or with VISTA have chosen to divide the sky into a grid of tracts across the sky, with each tract further broken up into smaller patches \citep{Aihara2018}. However, in regions close to the celestial pole, this becomes problematic without a coordinate transformation, i.e., Celestial to Galactic, resulting in a heterogeneous data product. Here, in order
to assess coverage and cadence, we have chosen to use the HEALPix scheme \citep{Gorski}. A strong motivator in this case is widespread support. HEALPix maps each pixel uniquely to a sky location based on its array index. 
%- no world coordinate system is required, although in cases 
Furthermore, for sparsely populated HEALPix maps a second array may hold the HEALPix index number. In HEALPix, the parameter NSIDE describes the total number of pixels in the all-sky tessellation, which corresponds to 12$\times$NSIDE$^{2}$. The angle subtended by a HEALPix pixel scales inversely with NSIDE.
For example, an NSIDE=1024 HEALPix map has pixels that subtend 3.4$\arcmin$ on a side while an NSIDE=2048 map has pixels that subtend 1.7$\arcmin$ on a side. Since survey characteristics would typically need to be defined over a wide survey area but sampled spatially by the instantaneous field of view of the survey, the HEALPix representation is compelling for storing cadence maps, sensitivity information and duration of survey.

%Except for cosmic microwave background imaging, 
Astronomical imaging data is typically projected onto HEALPix pixels for all sky representations of the data. However, a peculiarity of HEALPix is that while all pixels are of equal area, they are not equally shaped, nor are they rectangular (they are actually parallelograms). As a result, making the annotated coadds will generally require reprojection of the parent datasets into the HEALPix basis. Furthermore, in general the HEALPix pixels are much larger than typical pixels for many imaging datasets. For example, for HEALPix NSIDE=8192, a commonly adopted layout, %(the largest commonly supported value in term of number of pixels) the typical 
pixel size is roughly half an arcminute across, as opposed to many ground and space-based instruments which have pixel sizes measured in arcseconds. Finally, HEALPix pixelization is on a curved surface while astronomical imaging data is typically stored in tangent-plane projections of the sphere.

\begin{figure}
    \centering
    \includegraphics[height=1.5in]{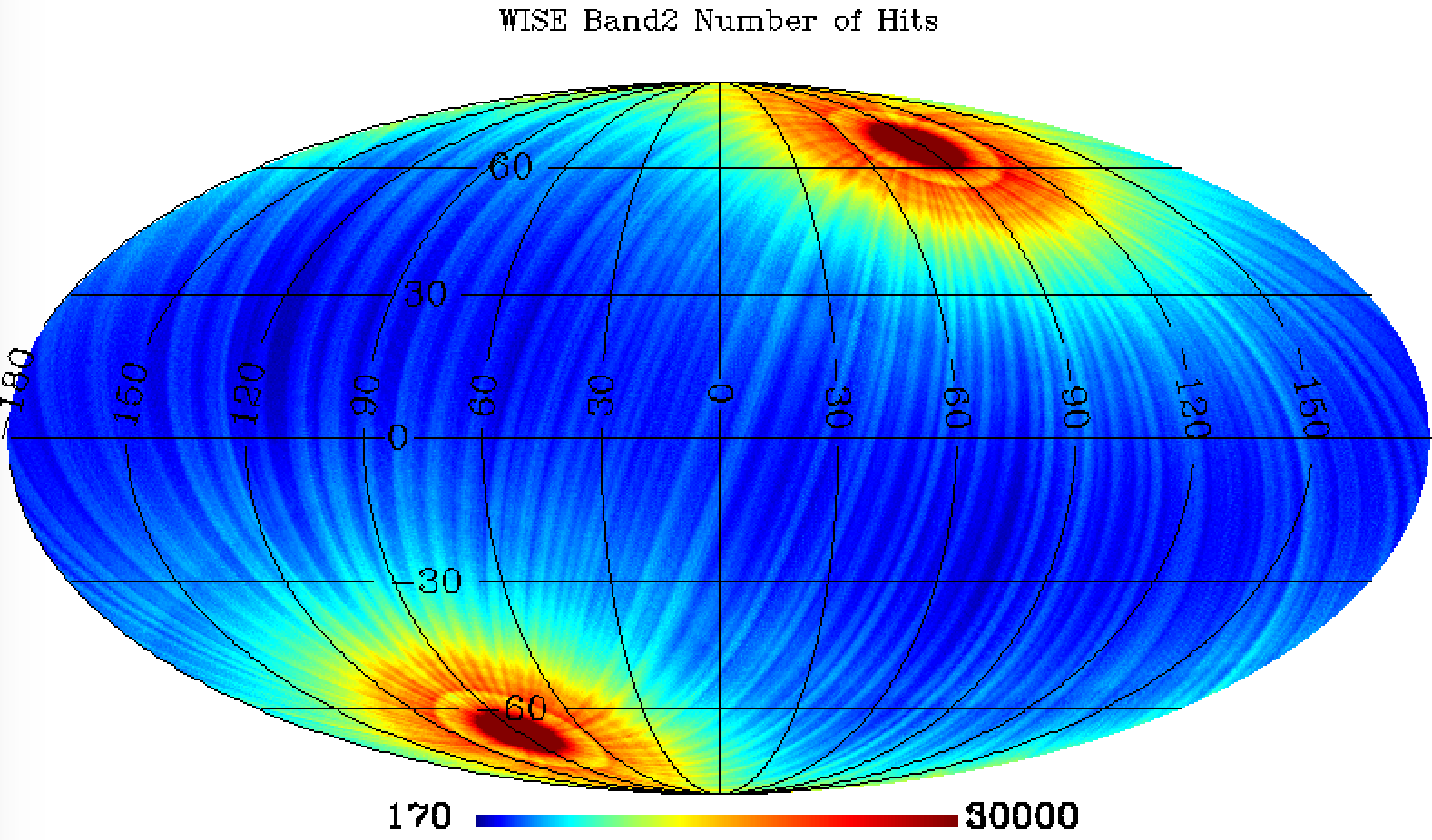}
    \includegraphics[height=1.5in]{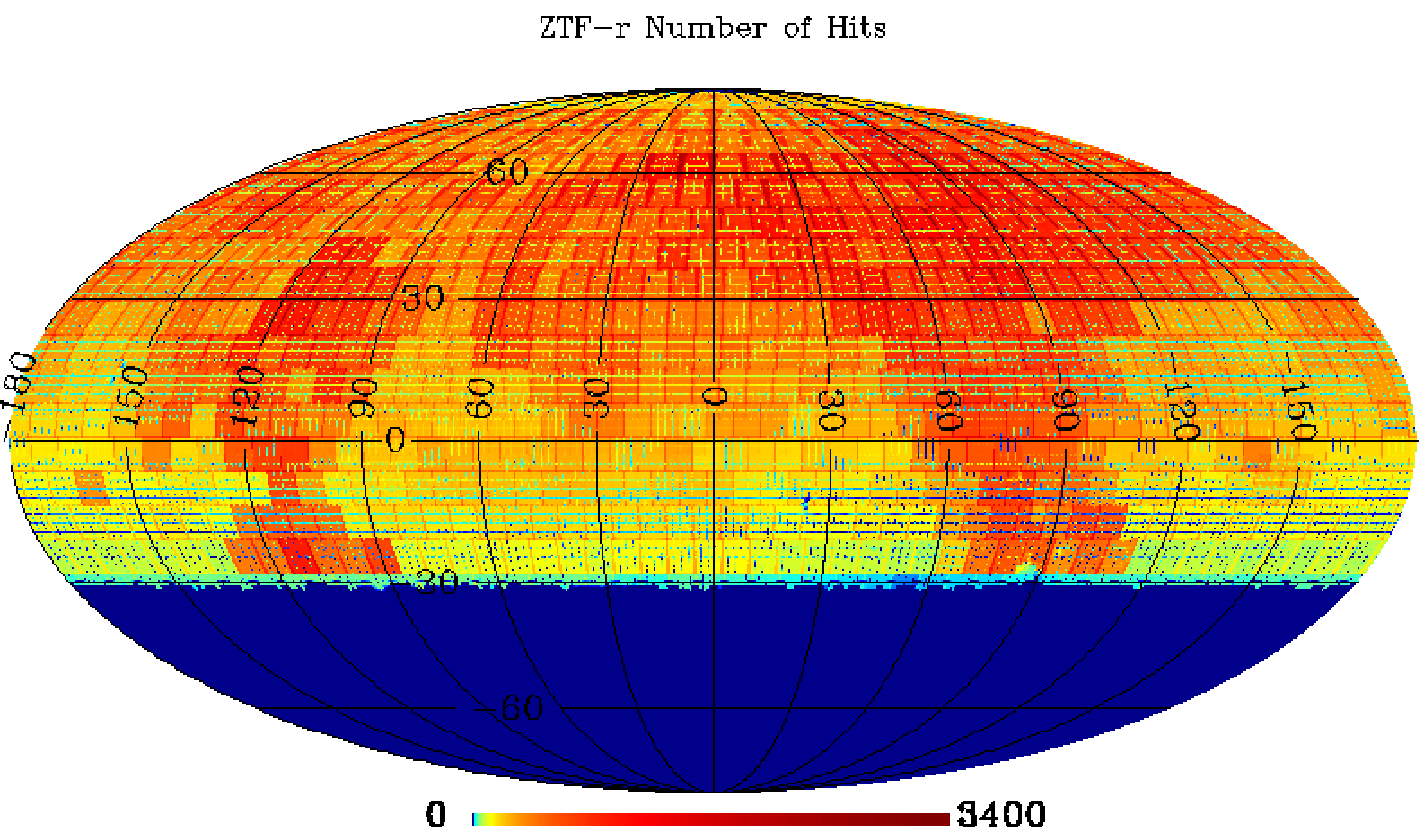}
    \caption{The left panel shows an all-sky HEALPix map in celestial coordinates of the number of frames taken by the {\it WISE} survey at 4.5$\mu$m. Due to the sun-synchronous orbit of {\it WISE}, the ecliptic poles are seen two orders of magnitude more times that regions near the ecliptic plane. For comparison, the right panel shows the number of frames taken by ZTF in the r-band. This shows that the southern sky was inaccessible with ZTF while the northern sky had relatively homogeneous coverage. In conjunction with additional planes that could illustrate the duration of the two surveys, one can assess in which regions of sky one can search for simultaneous variability in WISE and ZTF. The figure is therefore used to illustrate the value of storing survey characteristics in searches for variable phenomena.}
    \label{fig:surveyhits}
\end{figure}

Once survey characteristics are available, as shown in Figure \ref{fig:surveyhits} for two different surveys, identification of a time-variable optical/near-infrared source requires data on near-native pixels. 
The {\it algorithms} in the following sections that describe computing the actual variability metrics are not intrinsically tied to any one specific tessellation or tiling scheme. To place them onto the HEALPix basis however, care must be taken not to dilute any variability signal. For example, the actual data value assigned to the HEALPix pixel might be the maximum variability metric value in any data pixel that overlaps the HEALPix pixel. In a typical use case, the metrics estimated in HEALPix pixels would then be assessed to identify a region of sky of potential interest. The end user could then download the parent dataset of interest or more efficiently, just the stored light curves, for a more detailed analysis.

\section{Characterization of  Survey Cadence} \label{sec:cadence}

It is difficult to interpret the annotated coadd variability metrics for a survey without first understanding the underlying basic temporal data characteristics (e.g. first and last observation date, spacing between observations, sensitivity per epoch, etc). An exploration of data will usually begin by first examining whether the relevant data even possesses a temporal history suitable for the questions being asked. The cadence characterization encodes time information regarding the specific observation times onto the HEALPix spatial basis. We first demonstrate this using the WISE and {\it Spitzer} data sets.

\subsection{WISE and {\it Spitzer}}

The WISE mission was launched in 2009 and has been surveying the sky with a highly repetitive observation pattern which maps out the entire sky every 6 months \citep{Cutri}. Regions close to the ecliptic plane are thereby observed in 6 month intervals while regions close to the ecliptic poles have near continuous coverage. In this work, we focus on the 3.4 and 4.6$\mu$m bands of WISE since those wavelengths have coverage spanning several years, unlike the two cryogenic bands at 12 and 22$\mu$m.
The spatial resolution of WISE at the two shorter wavelength bands is 5.7$\arcsec$ with an instantaneous field of view of 47$\arcmin\times47\arcmin$.

In comparison, {\it Spitzer} operated as an observatory with community-defined targets and depths.
The {\it Spitzer}/IRAC instrument \citep{Fazio} only had a field of view of 5$\arcmin\times5\arcmin$ but the spatial resolution was better, with  a FWHM of $\sim$2$\arcsec$ at 3.6 and 4.5$\mu$m. The coverage of any particular region on the sky is therefore highly heterogeneous, except for the IRAC Calibration Field described later \citep{Krick2008}.

\subsection{Preprocessing of data}
\label{subsec:preprocess}

For our initial investigation of metrics, we prepared
small (typically less than ten arcminutes on a side)
fields from NEOWISE data in band W1 (3.4 microns) and band
W2 (4.6 microns). We downloaded all images with quality
flags 5 or 10 with any pixels overlapping the target
area. Following the unWISE processing steps \citep{Lang2014},
all NaN pixels were replaced by the mean of the four
adjoining non-NaN pixels. Then, the SWarp program
\citep{bertin2010} was used to re-grid all images onto a common
tangent projection (North-up, East-left) with the native pixel size of 2.75$\arcsec$, using a Lanczos-3 interpolation kernel. This pixel size is well-matched to the WISE point spread function size. Pixels with bad data quality i.e. where bits
9, 10 and 19 only were set, were masked. The application of the
interpolation kernel expanded single masked pixels in the
input images to clumps of NaN pixels in the interpolated
images. These re-gridded images were finally sorted by
time order and assembled into a cube, and output to a FITS
file along with a table including the timestamp of each
image plane. The treatment of the ZTF data that were used in this work is 
described in Section \ref{subsec:csdrl}.

%Also shown is the month-averaged 5$\sigma$ sensitivity of the data which was calculated by scaling single-epoch sensitivities with the inverse square root of number of epochs per month. This is not strictly accurate since {\it Spitzer} adopted a range of exposure times per frame while WISE had a fixed exposure time.
%As a result, the scaling should account for the individual exposure times. However, in this exercise, the plots are meant to be representative.

\subsection{Coverage Plots: Time Range of Data} \label{subsec:calendar}

\begin{figure}
    \centering
    %\epsscale{0.35}
    \includegraphics[width=3.8in]{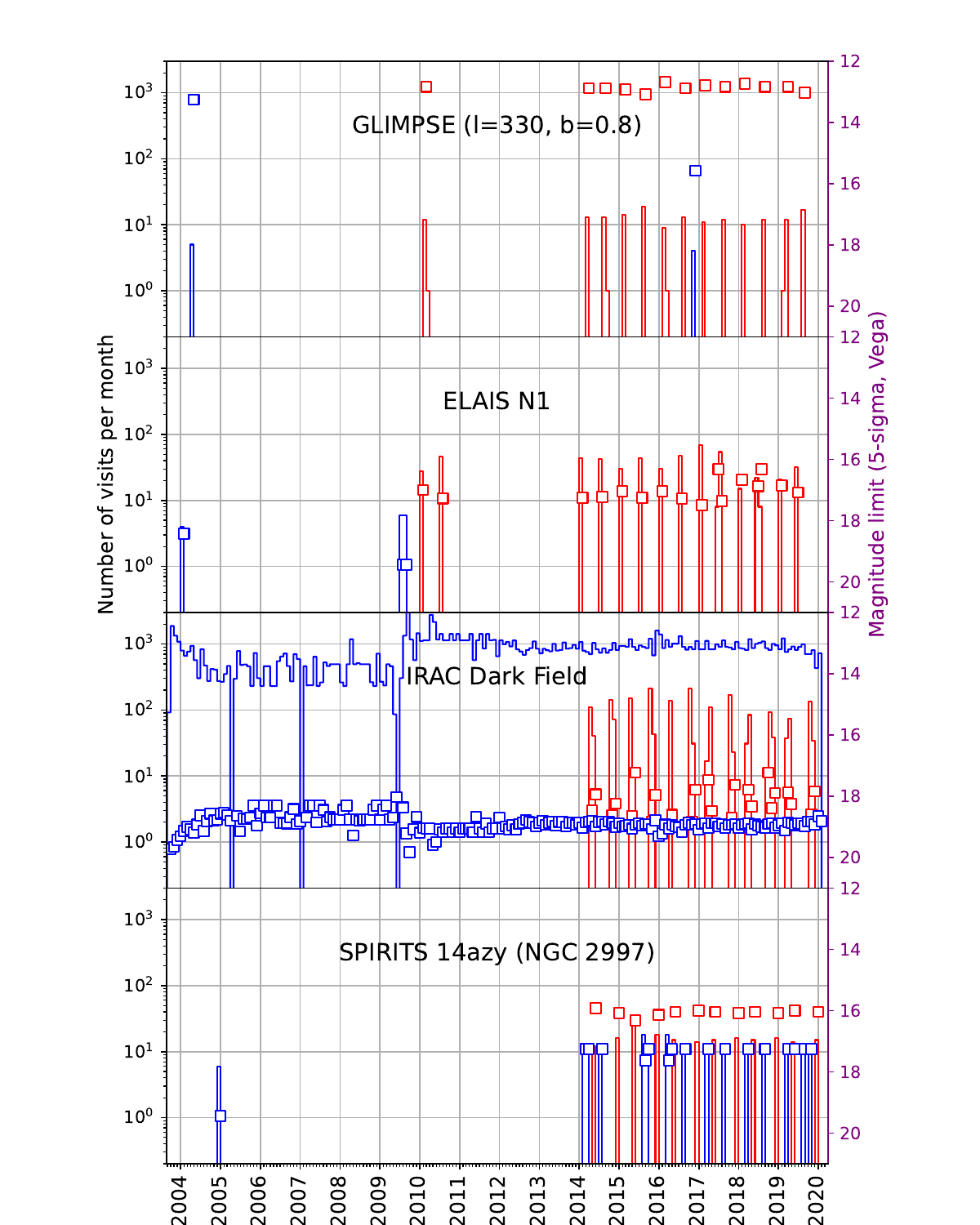}
    \caption{The plot shows surveys of the four different fields listed in Table 1 in four panels. The histogram of the number of visits per month as a function of calendar year corresponds to the values on the left vertical axis. The sensitivity of the monthly coadd resulting from those visits is shown as empty squares and corresponds to the values on the right vertical axis. {\it Spitzer} coverage is shown in blue while WISE coverage is shown in red. The IRAC Dark Calibration field was observed monthly over the {\it Spitzer} mission lifetime while the other fields had rather sparse {\it Spitzer} coverage. However, WISE observed the other fields with a near 6-month cadence with the Galactic Plane (GLIMPSE) fields having fewer visits per month than a field like ELAIS-N1. This indicates that the IRAC Dark Field data is suitable for generating highly sampled, long-duration light curves of faint sources. In contrast, the GLIMPSE field will only allow a search for slowly varying bright transients while the ELAIS-N1 field will allow slowly-varying but fainter transients to be detected. The SPIRITS survey  targeted nearby bright galaxies to detect transients with {\it Spitzer}.}
    \label{fig:calendar}
\end{figure}

To illustrate the principle of the annotated coadds, we queried the {\it Spitzer} and WISE/NEOWISE database at the central coordinates of a few well-studied fields which have different cadence of observations (Table \ref{fielddescription}). 
The GLIMPSE field is in the Galactic Plane region \citep{Benjamin}. The ELAIS-N1 field is a low cirrus extragalactic field, while the SPIRITS survey \citep{SPIRITS2017} targeted nearby bright galaxies looking for infrared transients - NGC2997 is one such galaxy in that sample. The IRAC Dark Field \citep{Krick2008} is a {\it Spitzer} calibration field in the vicinity of the North Ecliptic Pole.
The time stamps of these four representative fields with both {\it Spitzer} and WISE $\sim$3.4$\mu$m and 4.6$\mu$m coverage, along with the number of visits per month are shown in Figure \ref{fig:calendar}. 
We note that although Figure \ref{fig:calendar} shows the {\it Spitzer} data, we do not use those images in calculating variability metrics in the forthcoming sections due to the significant PSF mismatch between the {\it Spitzer} and WISE data. 
Figure \ref{fig:calendar} also shows the nominal sensitivity per month for these fields. 

The sensitivities for WISE/NEOWISE-R are based on a sqrt(N) 
scaling of single-frame sensitivity,
while the {\it Spitzer} sensitivities
are scaled from the SENS-PET exposure time calculator tool.
For a compact representation in Figure \ref{fig:calendar},
the sensitivities are shown by month,
and the single-frame sensitivities can be recovered by scaling from the number of exposures per month. The annotated 
coadd product will contain sensitivity percentiles on a
single-frame basis.
As can be seen, since the IRAC dark field was a calibration field, it was observed almost continuously every month for 16 years by {\it Spitzer}. Correspondingly, the number of visits per year is very high. In contrast, the GLIMPSE field was only observed twice by {\it Spitzer}, once in $\sim$2004 and once in $\sim$2016. WISE on the other hand scans the entire sky every six months with the ecliptic poles visited more often than the Galactic Plane. As a result, the IRAC Dark Field has many more WISE visits per month with correspondingly deeper sensitivity, while the GLIMPSE field only
has $\sim$10 visits per month and is relatively shallow.

\subsection{Cadence Plots: Histogram of time intervals} \label{subsec:histograms}

\begin{figure}
    \centering
    %\epsscale{0.35}
    \includegraphics[width=3.8in]{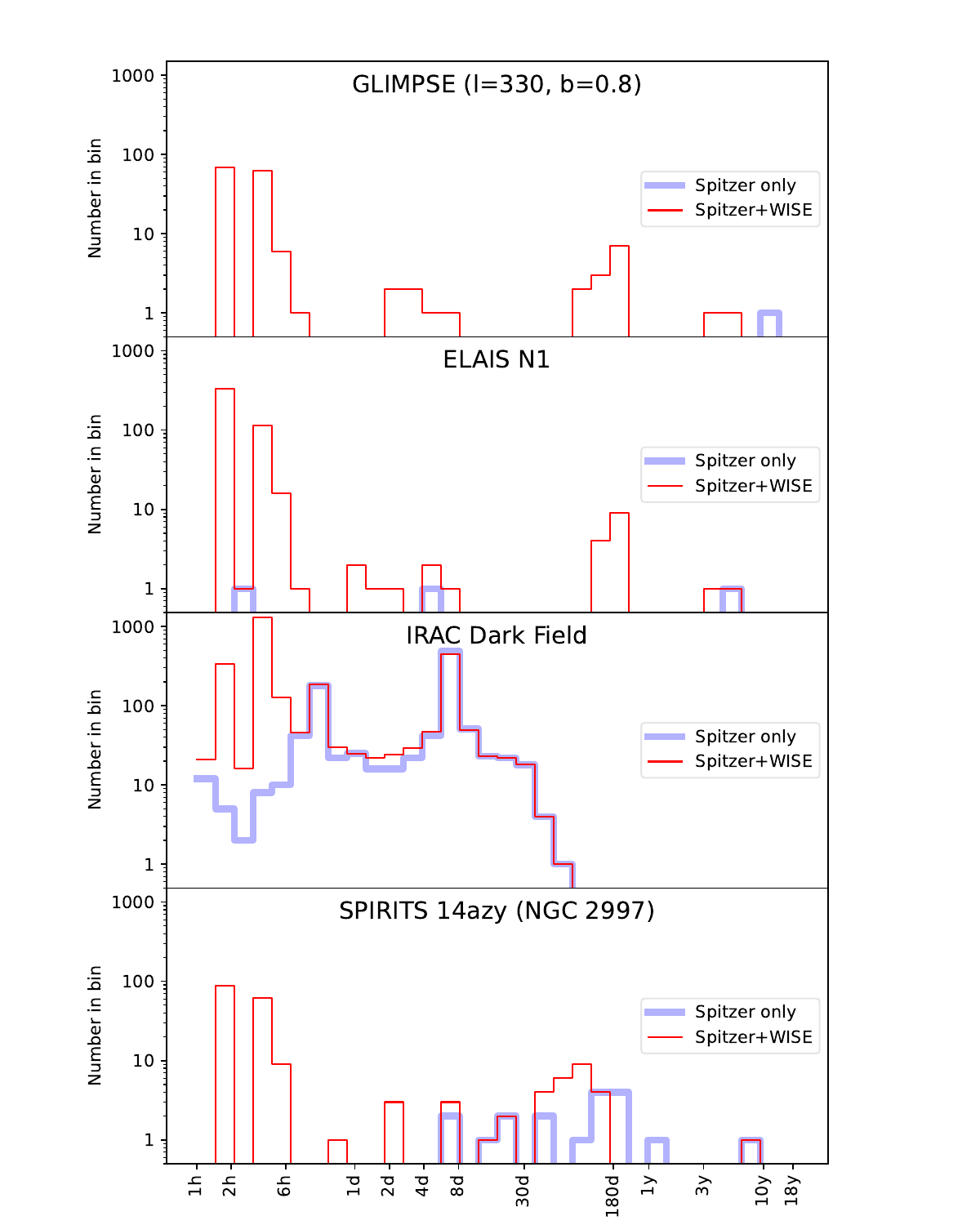}
    \caption{Plot showing the sampling cadence of {\it Spitzer} and WISE observations in four different fields obtained by taking the histogram of the time stamp difference between adjacent frames, as described in the text. Fields like the IRAC Dark Field at the North Ecliptic Pole which are observed every orbit by WISE and every month for calibration reasons by {\it Spitzer}, have time histograms extending from 1 hour up to 40 days. The other fields are observed more intermittently. For comparison, variable stars have periods of a few hours, supernovae show brightness evolution on timescales of days while AGN typically show variability on timescales of months.}
    \label{fig:successive_hist}
\end{figure}

In addition to the temporal extent over which the observations target the field, for variability analysis, it is also useful to know how often or with what typical frequency a field was observed. This is generated by taking the time stamps of all the observations sampled and generating a histogram of the differences between the times. 
Figure \ref{fig:successive_hist} shows cadence histograms for {\it Spitzer} only
and {\it Spitzer} and WISE combined for several fields.
%, where cadence is defined
%as the successive time differences of visits at the same %pixel.

%\subsection{Example Applications} 
\label{subsec:plots_application}

%In the following two examples we have computed cadence maps. The procedure was the same for both datasets. 
The time of observation, as well as the location of the four corners of each detector on the sky, were retrieved for all exposures currently in the IRSA archive. The observations are filtered by wavelength, and then time-ordered. For each observation, the four corners are used to compute the intersection of two (sets of) polygons: the detector footprint and the HEALPix pixels. For every HEALPix pixel that is touched by the detector footprint, the time of observation is added to a list associated with that pixel, ultimately resulting in a list of observation dates associated with every pixel on the sky. 
%Several maps are immediately derivable from these observation date lists. Maps of the first and last time of observation are the start and endpoints of each list. 
The total depth of coverage is simply the number of elements in each list assuming a constant single exposure time for the individual epochs.

The lists of observation dates per pixel are then converted to frequency histograms of the deltas between successive exposures. The histogram bin centers are spaced logarithmically in units of hours. The 30 bin centers span a time range of roughly 7.5 seconds to three years. The use of the logarithmic bins results in a compact storage format. In most use cases 16-bit integer representations for the bins (a total of 65535 observations per pixel) will suffice, resulting (for NSIDE=512, or 3.1 million pixels) a total file size of order 200 MB, easily within current computing capabilities. For datasets with depths of coverage exceeding this level, the use of fixed point notation combined with a normalization factor will allow larger values. It should also be noticed that in general these files can be losslessly compressed (i.e. by gzip) by a factor of ten or more. As a result, file data transfer is inexpensive. However, extending this to NSIDE=8192 (the largest commonly supported value) would result in a 50GB filesize, which is unduly cumbersome even when combined with data compression. In this case, we further quantize the allowed bin values. No IEEE standard exists for numerical storage using less than eight bits, but we can pack the bit storage to allow an effective 4-bits, or 16 discrete histogram levels. A simple lookup table can be used to map these 16 values into whatever values are required. In this fashion, the storage requirements can be decreased by an additional factor of four.

The data are stored in a Hierarchical Data Format (HDF5) file structure. This stores the histograms as a two-dimensional data array where one dimension is the HEALPix index, and the other is the 30 time-bin values. The HDF5 container additionally stores the bin centers and edge values. The HDF5 file format is widely supported. We additionally have converted the files for ease of use to IDL\footnote{Interactive Data Language; \url{https://www.l3harrisgeospatial.com/Software-Technology/IDL}} compressed savesets, as well as a series of FITS\footnote{Flexible Image Transport System; \url{https://fits.gsfc.nasa.gov}} files, where each file contains an all-sky map for a specific histogram bin.

\begin{deluxetable}{cccc}
\label{fielddescription}
\tablecaption{Fields Considered in this Study}
\tablehead{
\colhead{Name} & \colhead{R.A.} & \colhead{Dec} & \colhead{Area}\\
\colhead{} & \multicolumn{2}{c}{(J2000)} & \colhead{deg$^{2}$}
}
\startdata
GLIMPSE & 16h02m15.13s & -52d02m56.69s & 128\\
ELAIS-N1 & 16h11m26.27s & 54d36m23.12s & 9.4 \\
IRAC Dark Field & 17h40m05.39s & 68d59m16.56s & 0.06\\
NGC2997 & 09h45m38.95s & -31d11m29.35s & 0.04\\
\enddata
\end{deluxetable}

\section{Pixel-Based Variability Metrics} \label{sec:metrics}

We take the pre-processed image data as described in Section~\ref{subsec:preprocess} and compute
some variability metrics. A listing of the sources we analyzed, the survey and
wavelengths used, and the type of metric computed is shown in Table~\ref{srclist}.

\begin{deluxetable}{ccccc}
\label{srclist}
\tabletypesize{\scriptsize}
\tablecaption{Variable Sources Analyzed}
\tablehead{
\colhead{Name} & \colhead{R.A.} & \colhead{Dec} & \colhead{Metric} & \colhead{Source Type, Band, Figure}
}
\startdata
EP And & 01h42m29.34s & +44d45m42.45s & Reduced $\chi^{2}$ & Eclipsing binary; W2; Figure \ref{fig:chisq-epand}\\
WISEA J204027.30+695924.1 & 20h40m27.30s & +69d59m24.1s & Reduced $\chi^{2}$ & High proper motion; W2; Figure \ref{fig:chisq-j2040}\\
M51 Field & 13h29m56s & +47d13m47s & Reduced $\chi^{2}$ & Different variables; W2; Figure \ref{fig:M51}\\
SN 2017eaw & 20h34m44.24s & +60d11m35.84s & Consecutive signed slopes & Type IIp SN; W1; Figure \ref{fig:2017eaw-lc},\ref{fig:2017eaw-w1}\\
SN 2018cfa & 16h49m39.11s & +45d29m32.24s & Consecutive signed slopes & Type Ia SN; ZTF r-band; Figure \ref{fig:2018cfa-zr}\\
AV Men & 06h19m07.85s & -78d35m08.50s & Consecutive signed slopes & RR Lyrae-type star; W1 \& W2; Figure \ref{fig:AVMen-w1} \\
$\cdots$ & $\cdots$ & $\cdots$ & Tensor Decomposition & Figure \ref{fig:tensor}\\
\enddata
\end{deluxetable}

\subsection{Reduced chi-squared metric} \label{subsec:chi_squared}

The reduced chi-squared metric for variability requires two ingredients: a model which is
being tested against, and a measure of the variance that is needed to assess
the significance of deviations from the model value. Our model is simply the
median value ($I_{median,i}$) in each sky pixel ($p_{it}$ where $i$ is the spatial index of the pixel and $t$ is the time epoch).
For the variance, we computed a robust sigma
image from half the difference of the 84.13 percentile and the 15.86 percentile,
which for a Gaussian distribution gives the value of sigma. Our variance model
was simply a linear fit to:
\begin{equation} \label{eqn:sig}
\sigma^2_i = x_0 + x_1 * I_{median,i}
\end{equation}
using a robust least-squares fit with a softening parameter, from the
SciPy package \citep{2020SciPy-NMeth}. $x_{0}$ and $x_{1}$ are the linear polynomical coefficients used to derive the variance model.
This model was applied to the values
in our data cube to estimate a variance for each sky pixel at location $i$.
Then the chi-squared was computed as:
\begin{equation} \label{eqn:chi1}
\chi^2_i = \frac{1}{\sigma^2_i}\sum_{t} {{(p_{it} - I_{median,i})^2}}
\end{equation}
over all valid data values in the ``temporal stack'' ($t = 1...N$) for each sky pixel $i$.
The reduced chi-squared is then:
\begin{equation}
    \chi_{red,i}^2 = {\chi^2_i \over (N -1)}
\end{equation}
where $N$ is the number of valid data values per pixel $i$. We found that to screen out
artifacts, it was necessary to drop a small percentage of the maximum and
minimum values in each stack before computing the final chi-squared. We
settled on a rule of thumb of dropping 2\% to 3\% of pixel values to produce
an optimal chi-squared map. This procedure does eliminate single-visit
astrophysical measurements from fast-moving astronomical sources, in addition
to artifacts. However, variations in sky background, say due to moonglow or
due to variations in the zodiacal dust intensity would uniformly affect all pixels in the field of view and just increase the variance. This in turn would increase the threshold to classify a source as variable.

\begin{figure}[ht!]
    \centering
    \epsscale{0.5}
    \includegraphics[angle=0, width=1\columnwidth]{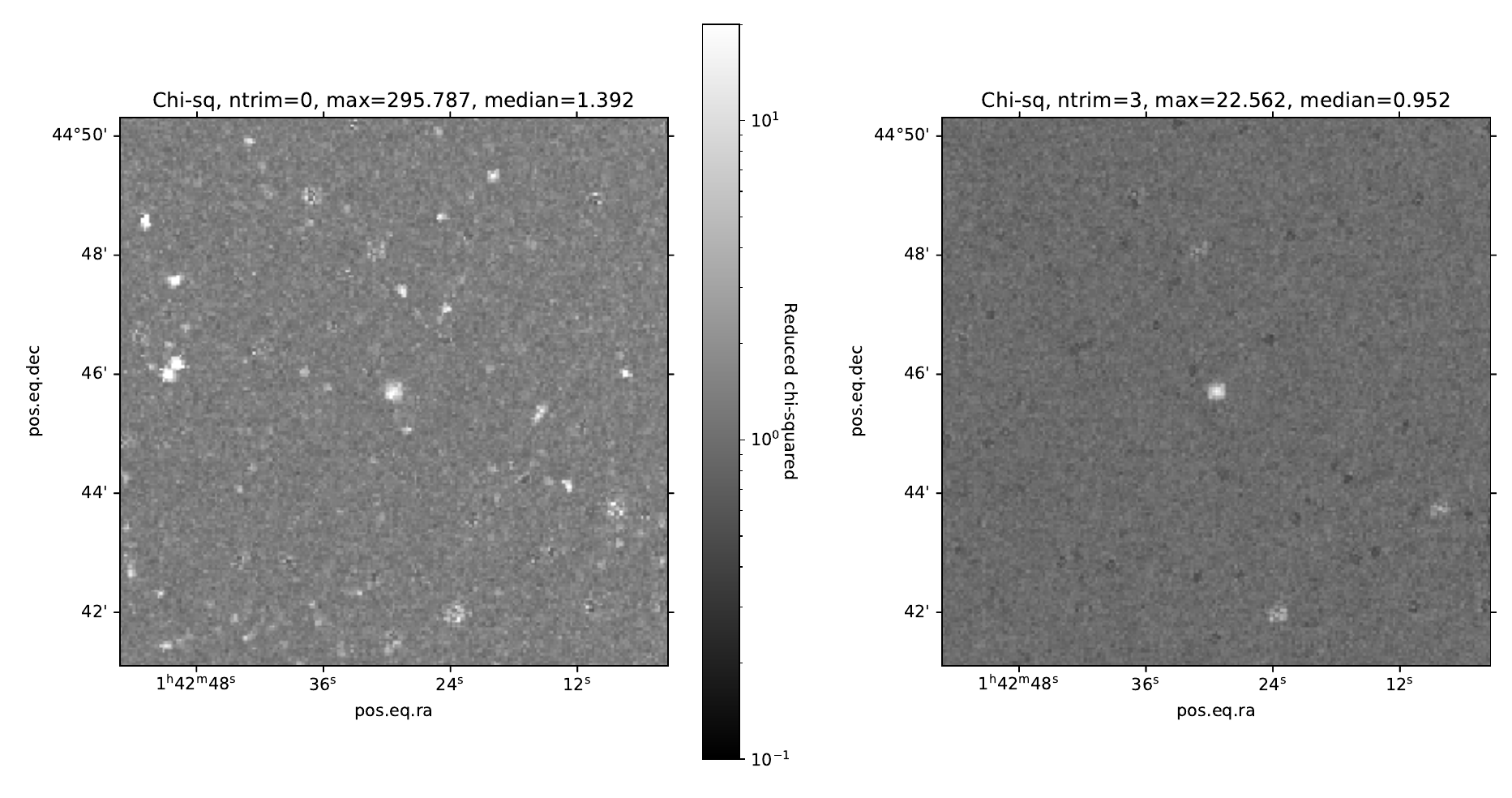}
    \caption{Reduced chi-squared images for EP\,And in NEOWISE W2 data. The left-hand side includes all data in the cube, and many artifacts are evident. The right-hand image excludes the three highest and lowest points in the time history of each pixel (corresponding to 2\%
    of images), removing artifacts but also possible fast-moving objects that would appear only in one frame. Most of the background pixels have reduced chi-squared values close to 1, indicating that our procedure for estimating the variance produces realistic values. At the image center, the reduced chi-squared values for EP And are shaped like the point-spread function, with a peak over 20. Bright stars in the field produce visible residuals in the chi-squared image, but the residuals appear noisy and do not follow the profile of a point source.}
    \label{fig:chisq-epand}
\end{figure}

\begin{figure}[ht!]
    \centering
    \epsscale{0.5}
    \includegraphics[angle=0, width=6.0in,trim={2cm 2cm 2cm  2cm}]{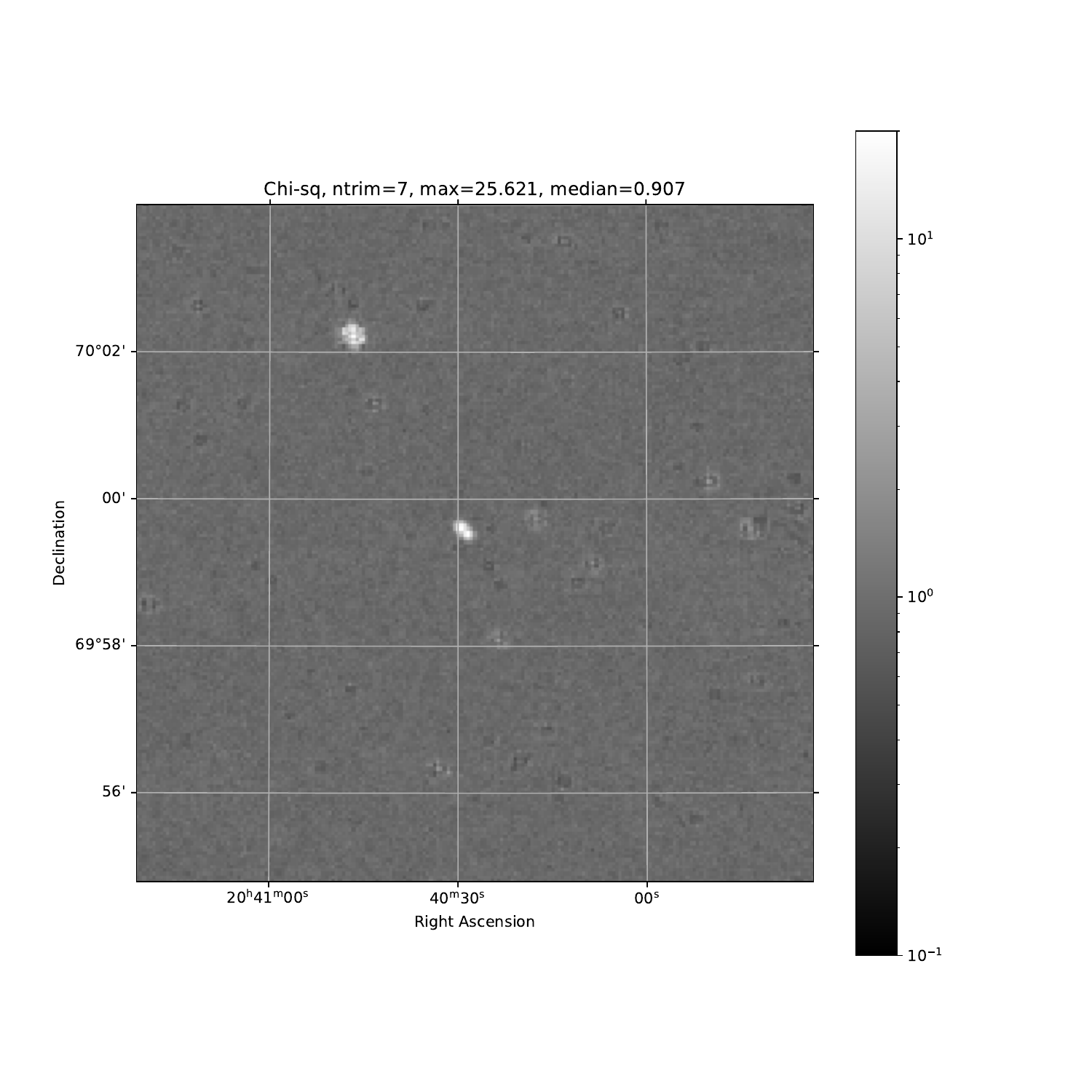}
    \caption{Reduced chi-squared image for the high proper-motion star WISEA J204027.30+695924.1 from NEOWISE W2 data. The image excludes the seven highest and lowest points in the time history of each pixel. This represent about 2\% of pixels at each point. Most of the background pixels have reduced chi-squared values close to 1, indicating that our procedure for estimating the variance produces realistic values. At the image center, the reduced chi-squared values show a double-lobed structure, with the largest value occurring where the wings of the point-spread function are uncovered. A bright star is visible to the northeast but is mottled and not shaped like the point spread function. The ntrim=7 is larger than in the previous case, because there are more image planes here than for Ep And.}
    \label{fig:chisq-j2040}
\end{figure}

By itself, a reduced chi-squared value is insufficient to assess the probability
of agreement with the median image. The degrees of freedom must be quoted along
with the reduced chi-squared. Formally, a $p$-value may be computed from the
reduced chi-squared and $N_{dof} = N-1$, where $p$ is the probability that
this value of reduced chi-squared, or a larger one, could arise by chance.
The important assumption to enable the mapping of a specific chi-squared value
to probability is that the underlying noise is Gaussian-distributed.

\subsubsection{Reduced chi-squared on variable stars}

The reduced chi-squared metric was applied to several
variable stars shown in Figure 15 of \cite{Masci2014}.
EP~And is a W~Uma-type eclipsing contact binary variable with a period of 0.404
days. The WISE W1 and W2 data consist of 222 frames between MJD of 56684 and 58337. The individual frames with 2.75$\arcsec$ pixels were aligned, background-subtracted and analyzed using the methodology described above.
The resultant $\chi^{2}$ images for EP And are shown in 
Figure~\ref{fig:chisq-epand}. 

As can be seen, executing the $\chi^{2}$ without trimming the outliers in the input frames results in a significant number of pixels with anomalously high $\chi^{2}$ values, which makes it challenging to cleanly identify the true variable source in the center of the image. However, trimming even a small fraction of frames, $\sim$1.5\%, as shown in the right panel of Figure~\ref{fig:chisq-epand}, results in the truly variable source in the center being much more clearly visible. The only other pixels which have significant $\chi^{2}$ values are due to some exceptionally bright sources in the field. However, they can be filtered out since they are significantly smaller in value and they do not have source brightness profiles consistent with that of a source.

\subsubsection{Reduced chi-squared on proper motion objects}

As another illustration of the $\chi^{2}$ metric on a source whose location varies rapidly, we selected the star WISEA J204027.30+695924.1.
J204027+69 was selected from a proper motion survey using WISE and the first year of NEOWISE data \citep{Schneider2016}, with high proper motion and a location near one of the ecliptic poles. This star has proper motions over 1.5 arcseconds per year in both right ascension and declination. 
The WISE data on this source consists of 552 frames between MJD of 56674 and 58325. We rejected a similar fraction $\sim$1.5\% of the frames, corresponding to 7 frames when creating the $\chi^{2}$ metric.
The chi-squared results are shown in Figure \ref{fig:chisq-j2040}. A double lobed structure is apparent, interpreted as the wings of the point-spread function being uncovered as the star's position changes. In comparison, the other pixels which have moderately large $\chi^{2}$ values are just residuals from the brightest stars. They can be eliminated either based on a threshold value or by filtering by the profile. 

\begin{figure}[h!]
    \centering
    \epsscale{0.1}
    \includegraphics[angle=0, width=\textwidth,trim={4cm 7cm 4cm  6cm}]{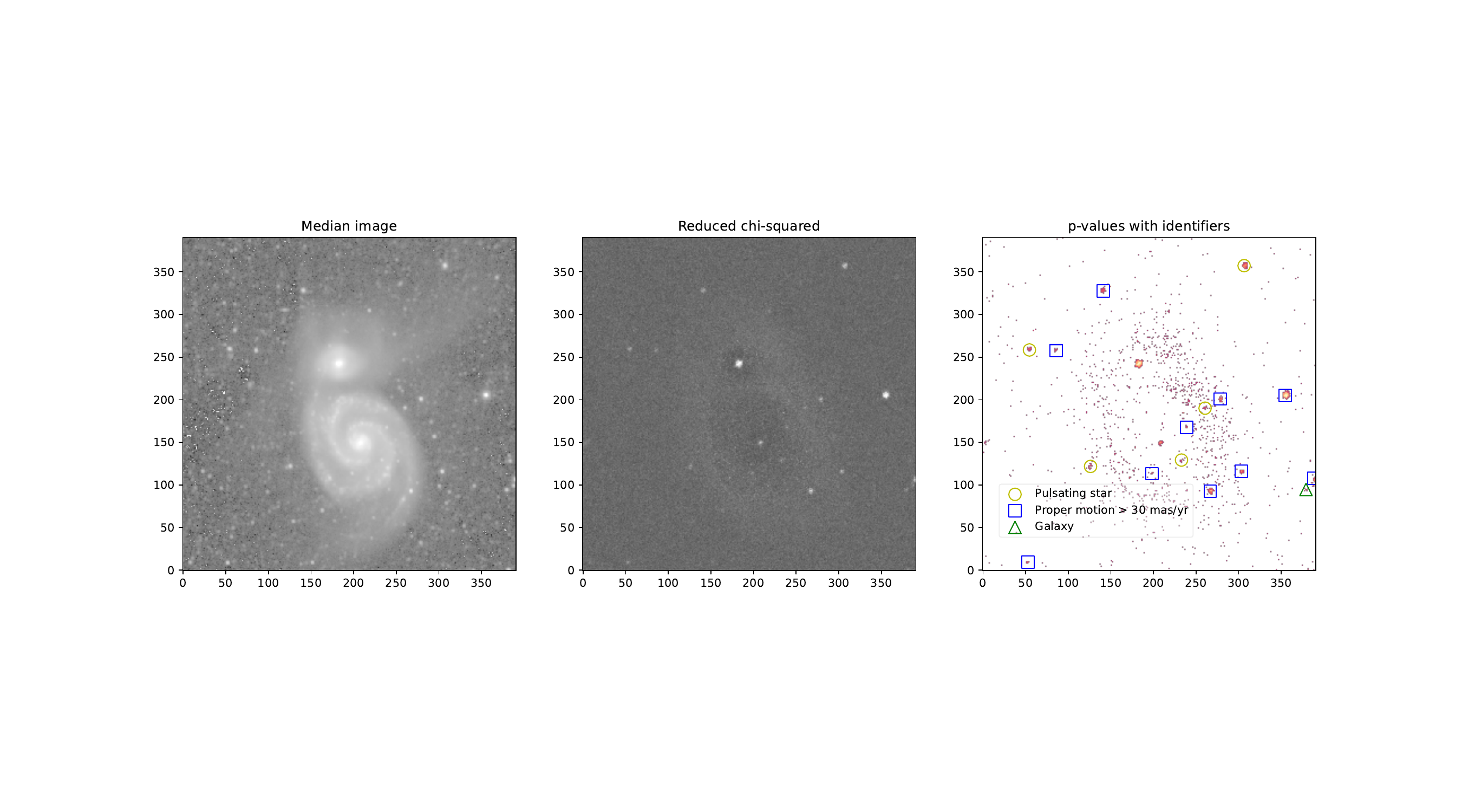}
    \caption{Left: Median image of M51 constructed from NEOWISE band W2 exposures. Middle: reduced chi-squared image with the three uppermost and lowermost values dropped. Right: reduced chi-square image thresholded so that only pixels with a p-value less than 0.001 are shown. Pulsating stars (empty circles), stars with proper motions (blue squares), and a background AGN 2XMM J132906.6 +470911 (green triangle) are identified. }
    \label{fig:M51}
\end{figure}

\subsubsection{Reduced chi-squared on M51 field}

The chi-squared metric was applied to a larger field (17.9 arcminutes on a side) containing the bright galaxy M51. Supernova 2011 dh occurred in this galaxy during the period when the WISE satellite was deactivated, and all traces of the event disappeared by the time of NEOWISE reactivation in 2014.
We analyze the 401 frames taken between MJD of 56644 and 58266. The median image and the reduced chi-squared image are shown in Figure \ref{fig:M51}. The lower right panel of that figure is thresholded for p-values below 0.001. We found that stars brighter than 17th magnitude in the Gaia G filter and with proper motions of 30 milliarcseconds per year or more are discernible in the chi-squared image. 9 out of 14 objects with a point-source-like spatial distribution fit these criteria. The remaining five Gaia stars are pulsators in the SIMBAD database.

These results show that the chi-squared metric is sensitive to stars moving by 150 milliarcseconds or just over 1/20th of a pixel over the 5-year NEOWISE baseline. Further, these results are achieved on top of the spatially-complex background of M51. 

Overall, we find that the $\chi^{2}$ metric results in a clear identification of variability. However, to minimize spurious detections, it is essential to trim the number of input frames, identify a threshold for the $\chi^{2}$ values based on simulations and use spatial filtering to reject out residuals from bright objects.

%---------------------Start of Frank's subsection------------------------------
%\newpage
\subsection{The Consecutive Signed-Difference Run Length (CSDRL) Metric}\label{subsec:csdrl}

We describe now a different metric for ``collapsing'' epochal image data and
detecting flux-variable candidates therein. This metric
complements the reduced chi-square metric (Section~\ref{subsec:chi_squared}) when
searching for various flavors of variability since it has its own benefits and
limitations that depend on the input cadence and timespan of a set of
observations. This metric is coined the Consecutive Signed-Difference Run
Length (CSDRL) value. An image consisting of CSDRL values is constructed
as follows:

\begin{enumerate}
  \item{A sequence of pre-calibrated image exposures (epochal images) are
        registered and interpolated onto a common pixel grid to construct an
        image cube. Pre-calibration includes any relative astrometric refinements,
        photometric-throughput (gain) matching, and background subtraction.
        There are two spatial dimensions and the third dimension is an epoch
        identifier (sorted by increasing observation time).}
  \item{Store the flux time series (lightcurve) for a single pixel in
        this grid from the sequence of time-ordered images.}
  \item{Apply a median filter to the pixel time series to suppress possible
        noise between the consecutive time-points. The median filter window
        length is an input parameter. Its choice depends on the timescale of
        the variability sought and observation cadence of the imaging.
        The filter-length can therefore be fixed given prior knowledge
        of the survey cadence and science case(s) for which to optimize searches.}
  \item{Compute the pairwise first-differenced time series for this pixel:
        $D_{t} = f_{t+1} - f_{t}$ for all fluxes $f_{t}$ at epochs
        $t = 0, 1, 2, ...$.}
  \item{Find and store the {\it longest} run length (or count) of
        {\it consecutive} pairwise slopes that are either positive or
        negative, i.e., $D_{t} > 0$ or $D_{t} < 0$ respectively.}
  \item{Generate an image of the longest runs by ``collapsing'' all pixel-based
        time series according to steps 3 - 5. A positive run-length implies
        a span of epochs where the pixel flux is increasing while a negative
        run-length implies a span where where the flux is decreasing across
        the (median-filtered) timepoints.}
  \item{A 2-D spatial sum-filter is applied to the image from step 6.
        This accentuates neighboring spatially-correlated pixels whose fluxes
        are either all increasing or decreasing with time, i.e., are associated
        with potentially the same point-source signal (spanning the
        instrumental PSF if sufficiently sampled). This operation also
        suppresses (cancels) run-length values where spatially-adjacent
        pixels have opposing signs for their run-lengths. This behavior may
        occur on scales of order the PSF size due to for example
        position-dependent systematic errors following the image registration process
        in step 1 (i.e., related to an inaccurate distortion calibration), or
        asymmetric PSFs and a rotating instrument field-of-view
        across epochs. This sum-filtering also suppresses slow moving objects
        where pixel flux rises in the direction of travel (yielding positive
        runs) and decreases in the opposing direction (yielding negative runs).}
  \item{The pixel run-length values in the image from step 7 are replaced with
        their absolute value. This means all negative run-lengths become positive
        values. The reason is that we are interested in detecting clusters of
        pixels with either excess positive or negative run-lengths, regardless
        of temporal trends in flux. We now have an image called the CSDRL
        metric image.}    
\end{enumerate}

Detections of excess signals on local pixel scales in the CSDRL image
represent candidates for transient or variable sources, i.e., which may
have exhibited a systematic trend in flux anywhere in their time-ordered
sequence of epochal images. This image is now ready for source detection
where by ``source'', we mean local pixel maxima in the CSDRL image
above some specific threshold and satisfying other criteria (see below).

First, the CSDRL pixel threshold is selected according to some statistical
significance (chance $P$-value) that is empirically calibrated from the local
CSDRL pixel distribution. Spurious fluctuations in the CSDRL pixel values
do not follow any standard parametric distribution (e.g., Gaussian)
and hence their frequency of occurrence is non-trivial to quantify in a probabilistic sense.
This is because first, the values are positive-definite (with negative run-lengths
wrapped to values $> 0$) and where runs on noisy pixels are more likely to follow a
Bernoulli process if and only if fluctuations are independent across
epochs, i.e., similar to tossing a fair coin. Another reason is because of systematic
errors associated with the imaging instrument, image pre-calibration,
and/or image-reconstruction process. For example, asymmetric PSFs and field-of-view
rotations; temporally-varying and/or spatially-dependent PSFs; inaccurate
calibration of field-distortion or motion aberrations leading to imperfect
image registration; image interpolation or resampling errors (most severe for
undersampled images); flux calibration errors or uncorrected drifts in
photometric throughput. Other biases intrinsic to the images such as intra-pixel
responsivity or temporally-varying complex backgrounds that are challenging to
subtract from each epochal image will also contribute.
Hence, the ``null hypothesis'' for calibrating $P$-values must account
for all local systematics that will impact the static (non-variable) source
population in the field from which the CSDRL image is computed. We call these
``local systematics'' since it is expected that each field's inputs will
behave differently, even if selected from the same homogeneous survey.

The single-pixel-based CSDRL $P$-values are computed from the pixel histogram.
The CSDRL threshold used to search for local maxima in the CSDRL image is based
on the quantile corresponding to one's choice of $P$-value or tolerance
for fraction of false positives (e.g., $< 0.1$\%). This choice is more
important if one were conducting a
blind search for flux-variable candidates. The detection process is as follows:
if a CSDRL pixel value exceeds the CSDRL threshold and that of its neighbors
centered on a $5\times5$ pixel region, and if 4 neighbors also exceed
the threshold within a $3\times3$ pixel region, the cluster is added to the
detections list. Additional metadata are then assigned, e.g., the sky coordinates
at the center of the cluster peak pixel, it's significance ($P$-value), and sum of all
CSDRL values within the local $3\times3$ region. A separate table consisting of
$P$-values as a function of the sum of local clustered CSDRL values is also
generated by re-executing the detector with an input threshold of zero.
This ancillary table is used to infer an approximate upper bound for
the $P$-value for a detected cluster of CSDRL values.

When a source has brief periods of variability, but is mostly quiescent, the reduced chi-square metric (Section~\ref{subsec:chi_squared}) dilutes the significance of variability. In contrast, the CSDRL
metric is relatively immune to the presence of additional epochs when a source is constant in brightness but that may
contribute pure random noise to the pixel-flux time series. This is because this
metric always searches for the {\it longest} run in consecutive pairwise flux
differences of the same sign. Once the longest run is encountered on a real
flux variable, it is stored until a longer run is encountered at later epochs
in the time series. Whether a longer run is found depends on the nature of
the flux variability, e.g., re-occurring variables (see Section~\ref{csvarstar})
versus transient events localized within some time span. Depending on the pre-stored
run-length for a pixel, it is also possible for random noise fluctuations to
result in a longer run-length by chance, but the probability of this happening
rapidly decreases (almost exponentially) with increasing run-length. The key
point is that for epochal image data where the instrumental PSF is sufficiently
sampled, clustered (or spatially-correlated) excesses in CSDRL values will
greatly inflate the statistical significance of a variable candidate.
Accounting for such spatial correlations is a crucial element of the detection
process.

%------------------------------------------------------------------------------
\subsubsection{Application of the CSDRL Metric to a Supernova Lightcurve in WISE}

\begin{figure}
  \centering
  \epsscale{0.5}
  \includegraphics[angle=0, width=6.0in]{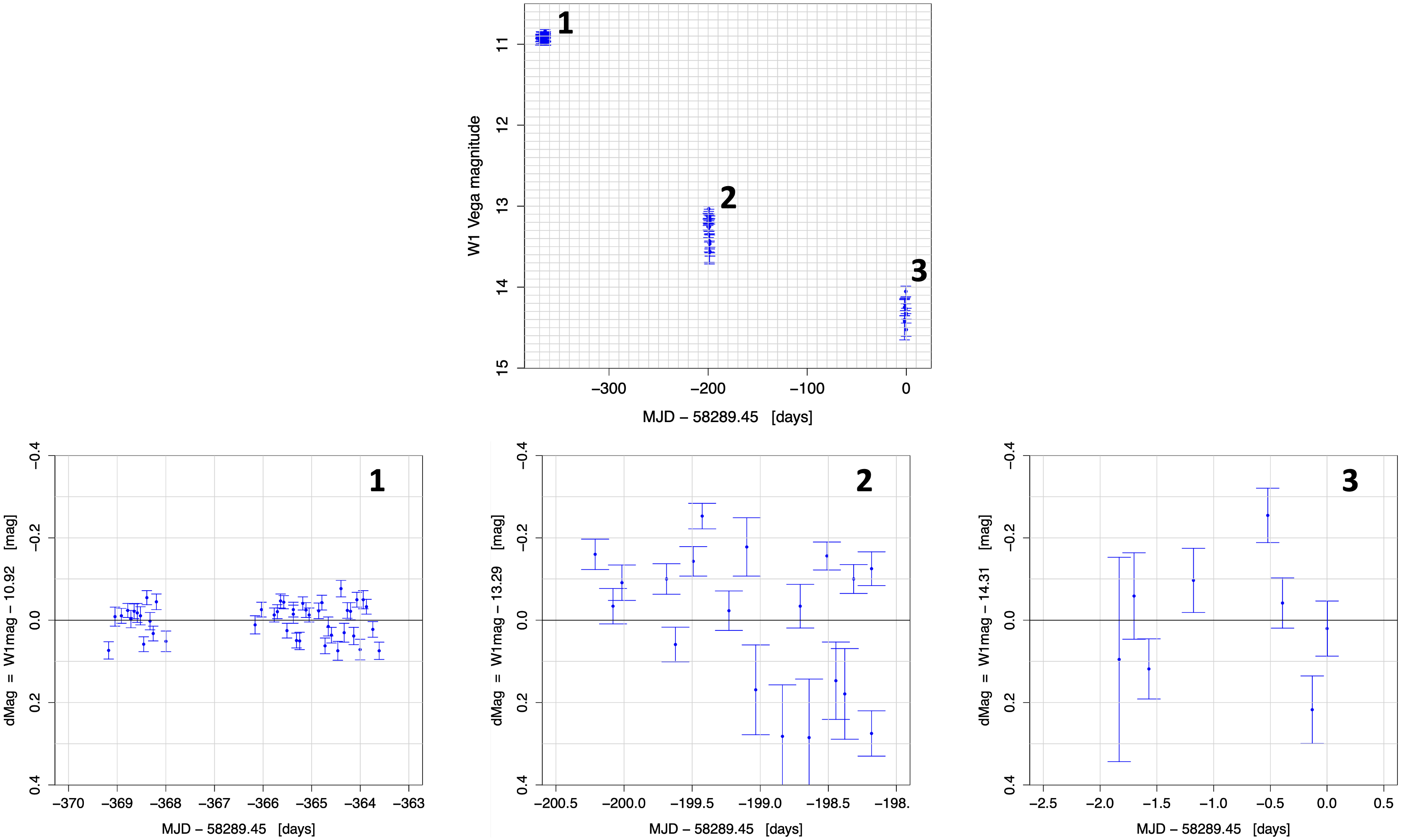}
  \caption{Lightcurve in the WISE W1 band for the Type-IIP supernova AT 2017eaw
           in NGC 6946. This supernova was detected in 72 good quality exposures
           from three coverage epochs spanning $\sim12$ months as WISE surveyed
           the sky. The three panels below are zooms on each sky-coverage epoch,
           plotted as a differential magnitude relative to the mean magnitude
           within each epoch.}
  \label{fig:2017eaw-lc}
\end{figure}

We tested the CSDRL method on a stack of 608 epochal images in the WISE W1
band centered on the Fireworks Galaxy, NGC 6946. This galaxy hosted a
Type-IIP supernova (AT 2017eaw) that was discovered on May 14, 2017.
During its continued surveying of the sky, the NEOWISE spacecraft happened
to scan over NGC 6946 in early June 2017, just after the supernova (SN) peak.
Due to the relatively slow decline in flux (not unusual for this SN class),
NEOWISE covered this region again and re-detected the SN in a subsequent pass
of the sky six months later (December 2017). This region was then covered
again six months later (June 2018) where the SN was just barely detected above
the W1 flux limit in NEOWISE's automated processing. The full lightcurve
is shown in Figure~\ref{fig:2017eaw-lc}. Only good-quality photometry was
retrieved from the NEOWISE archive, specifically the NEOWISE-R Single Exposure
Source Database (hosted by IRSA\footnote{\url{https://irsa.ipac.caltech.edu/Missions/wise.html}} at IPAC).

\begin{figure}
  \centering
  \epsscale{0.5}
  \includegraphics[angle=0, width=\textwidth]{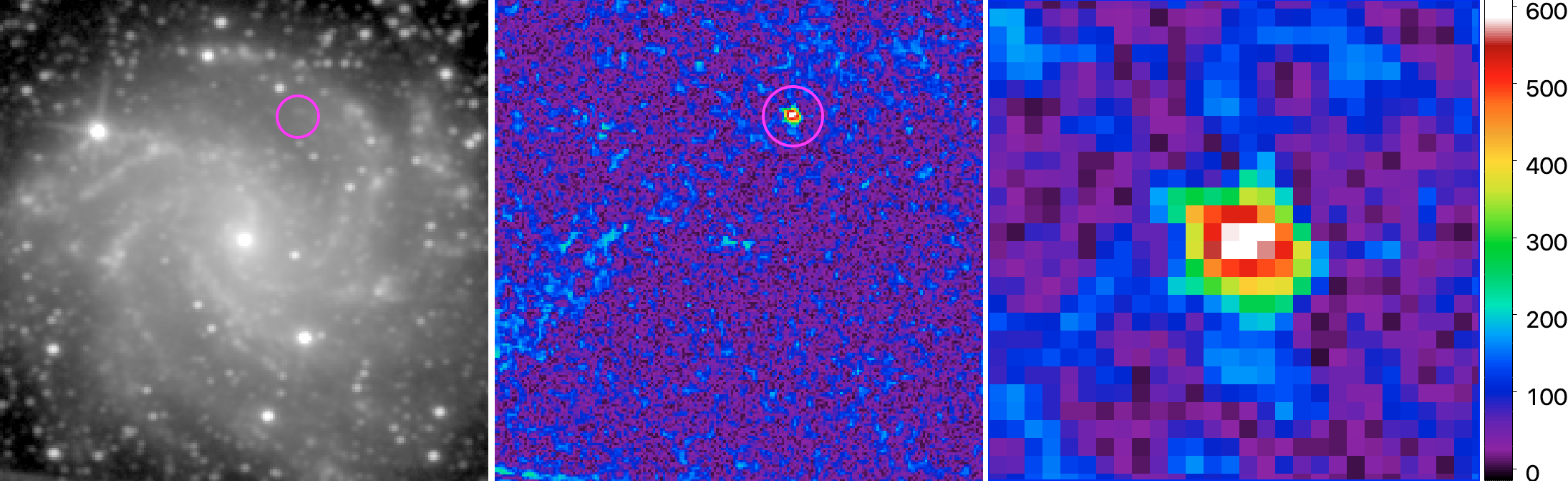}
  \caption{{\it Left}: median image that collapses 608 WISE W1 single-exposure images
           of NGC 6946 acquired in June 2017, with the purported position of
           SN 2017eaw circled in magenta (see Figure~\ref{fig:2017eaw-lc}
           for lightcurve). Image measures $\sim9.2\times 9.2$ arcmin$^2$.
           {\it Middle}: image of the CSDRL metric from collapsing all 608 exposures
           with the SN clearly detected.
           {\it Right}: zoom-in on SN 2017eaw on the CSDRL metric image
           measuring $\sim1.24\times 1.24$ arcmin$^2$. The color-bar represents
           the range in CSDRL pixel values.}
  \label{fig:2017eaw-w1}
\end{figure}

Following the recipe in Section~\ref{subsec:csdrl}, we applied a median
filter of length 51 timepoints to each pixel time-series from $\sim$ 608 time-ordered
epochal images before computing the CSDRL image.
This smoothing maximized the overall signal (run-lengths) in pixels
centered on the location of SN 2017eaw relative to the surrounding spurious
run-lengths resulting from noise fluctuations. The latter include the possible
systematics described above. The CSDRL image and a zoom-in on the SN
location are shown in Figure~\ref{fig:2017eaw-w1}. Pixels on the SN
peak within a $3\times3$ pixel region have CSDRL values of $\gtrsim 580$
compared to $\simeq 60$ for the background median value.

Based on the empirically-derived distribution of spurious CSDRL cluster
detections in this field, the probability of detecting the cluster of
CSDRL values at the position of SN 2017eaw by chance is $<\sim 0.07\%$.
This upper bound is simply $100/N\left(\sum{CSDRL}\,<\,maxthres\right)$ where
$N\left(\sum{CSDRL}\,<\,maxthres\right)$ is the number of spurious clusters with
summed CSDRL pixel values less than the threshold {\it maxthres}, here set to 4925,
-- the value measured for SN 2017eaw. This also happens to be
the maximum $\sum{CSDRL}$ value in this field.

\subsubsection{Application of the CSDRL Metric to a Supernova Lightcurve in ZTF}

We also tested the CSDRL method using image data from a ground-based survey
that covered another SN, in this case the ZTF survey. 
The Zwicky Transient Facility is a project specifically designed to search for time-variable phenomena \citep{2019PASP..131a8002B}. Initially motivated by SNe searches, the project has expanded over time to include a wide range of science cases, including variable stars, AGN, gravity wave electromagnetic counterparts, and solar system objects. ZTF surveys the sky with a large (47 square-degree) mosaicked sixteen CCD camera attached to the Palomar 48-inch telescope. The survey is normally carried out in three filters ({\it g},{\it r}, and {\it i}-band) using a highly irregular survey strategy which interleaves large-areal surveys with specific targeted observations (e.g. local galaxies) all of which possess different cadences (i.e. timescales on which they repeat). Due to this irregular observation strategy, it is impossible for most users to {\it a priori} predict which parts of the sky have been observed when. 

We produced a series of cadence maps using the HEALPix basis. These cadence maps were generated with NSIDE=512, which results in pixels roughly 7 arcminutes on a side. The choice of pixel size is such that the individual CCD detectors are sampled by 15 HEALPix pixels across, while the gaps between the chips are resolved. There are 3.1 million pixels total for this choice of NSIDE, but since ZTF only covers the northern hemisphere only 2.4 million are actually populated. The HEALPix arrays are stored sparsely by also storing an index array within the HDF5 file, which contains the explicit HEALPix value associated with array element. Because ZTF uses the same exposure time for all observations, there is a direct relation between number of exposures and sensitivity.

\begin{figure}
  \centering
  \epsscale{0.5}
  \includegraphics[angle=0, width=6in]{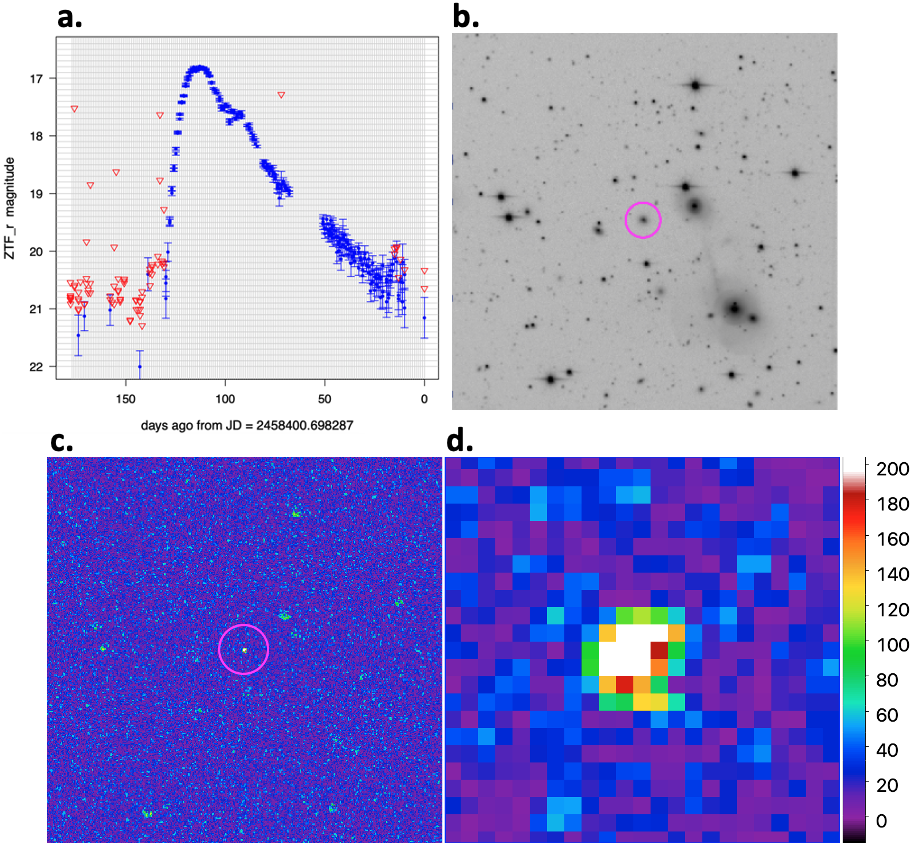}
  \caption{(a) Lightcurve of Type-Ia supernova 2018cfa discovered in the
           ZTF $r$-filter in June 2018. Blue points are $>3$-$\sigma$ detections
           and red triangles are 5-$\sigma$ upper limits (non-detections).      
           (b) Median image that collapses 335 ZTF $r$-filter exposures
           centered on the location of the SN. This appears to be associated
           with a marginally-resolved galaxy (WISEA J164939.30+452933.8;
           2MASS J16493924+4529335). Image measures $\sim8.1\times 8.1$ arcmin$^2$. 
           (c) image of the CSDRL metric from collapsing all 335 exposures with
           the SN clearly detected.
           (d) Zoom-in on SN 2018cfa on the CSDRL metric image measuring
           $\sim23\times 23$ arcsec$^2$. The color-bar represents the range in
           CSDRL pixel values.}
  \label{fig:2018cfa-zr}
\end{figure}

The image data from this
survey is subject to additional flavors of systematics not present in a
space-based survey such as WISE, for example, temporally varying PSFs
due to atmospheric turbulence and/or varying photometric throughput due to
changes in atmospheric transparency on short timescales. In particular, spatial
variations in transparency across individual exposure images can lead to uncorrected
residuals in instrumental throughput following photometric calibration.
ZTF discovered the Type-Ia SN 2018cfa on June 6, 2018. The SN signal was
detected in $\sim 300$ $r$-filter exposure images spanning $\sim 5.5$ months.
A lightcurve was generated using the ZTF forced-photometry
service \citep{ZTFforced}\footnote{\url{https://irsa.ipac.caltech.edu/data/ZTF/docs/ztf_forced_photometry.pdf}}. 
This is shown in Figure~\ref{fig:2018cfa-zr}a.

Following the recipe in Section~\ref{subsec:csdrl}, we applied a median
filter of length 11 timepoints to each pixel time-series from $\sim$ 335 time-ordered
epochal images before computing the CSDRL image.
This smoothing maximized the overall signal (run-lengths) in pixels
centered on the location of SN 2018cfa relative to the surrounding spurious
run-lengths resulting from noise fluctuations. The latter include all possible
systematics described above. The CSDRL image and a zoom-in on the SN
location are shown in Figures~\ref{fig:2018cfa-zr}c and d. Pixels on the SN
peak within a $3\times3$ pixel region have CSDRL values of $\gtrsim 195$
compared to $\simeq 17$ for the background median value.

Based on the empirically-derived distribution of spurious CSDRL cluster
detections in this field, the probability of detecting the cluster of
CSDRL values at the position of SN 2018cfa by chance is $<\sim 0.008\%$.
This upper bound is simply $100/N\left(\sum{CSDRL}\,<\,maxthres\right)$ where
$N\left(\sum{CSDRL}\,<\,maxthres\right)$ is the number of spurious clusters with
summed CSDRL pixel values less than the threshold {\it maxthres}, here set to 2293,
-- the value measured for SN 2018cfa. This also happens to be
the maximum $\sum{CSDRL}$ value in this field.

%------------------------------------------------------------------------------
\subsubsection{Application of the CSDRL metric to Variable Stars}\label{csvarstar}

Unlike transient events that monotonically increase and/or decrease in flux
within some timespan of an input sequence of epochs (e.g., the supernovae
above), the CSDRL method is also expected to detect runs in consecutive pairwise
slopes of the same sign falling between any of the peaks and troughs of a
re-occurring flux variable, whether periodic or aperiodic.
The only difference here is that since the {\it longest} runs
along each pixel's time series are retained when constructing the CSDRL metric,
pixels in clusters with excess CSDRL values may be out of phase with each
other, i.e., they may sample different sub-spans of the full input timespan.
This poses no issue since our primary goal is detecting where, in 2-D space,
spatially-correlated excesses in CSDRL value are located. Presumably, these excesses
correspond to activity originating from the same flux variable, regardless
of when it occurred. It's also worth noting that for re-occurring flux variables
in general, the significance of the CSDRL values are likely to improve as
the number of input epochs increases. This is because it is more likely
to encounter longer and longer runs in signed-slope between a peak and trough
that mitigate noise fluctuations. The CSDRL values will approach some maximum
that depends on the input noise, observation cadence, and characteristic
timescale of monotonic trends (rise/fall episodes).

\begin{figure}
  \centering
  \epsscale{0.5}
  \includegraphics[angle=0, width=6in]{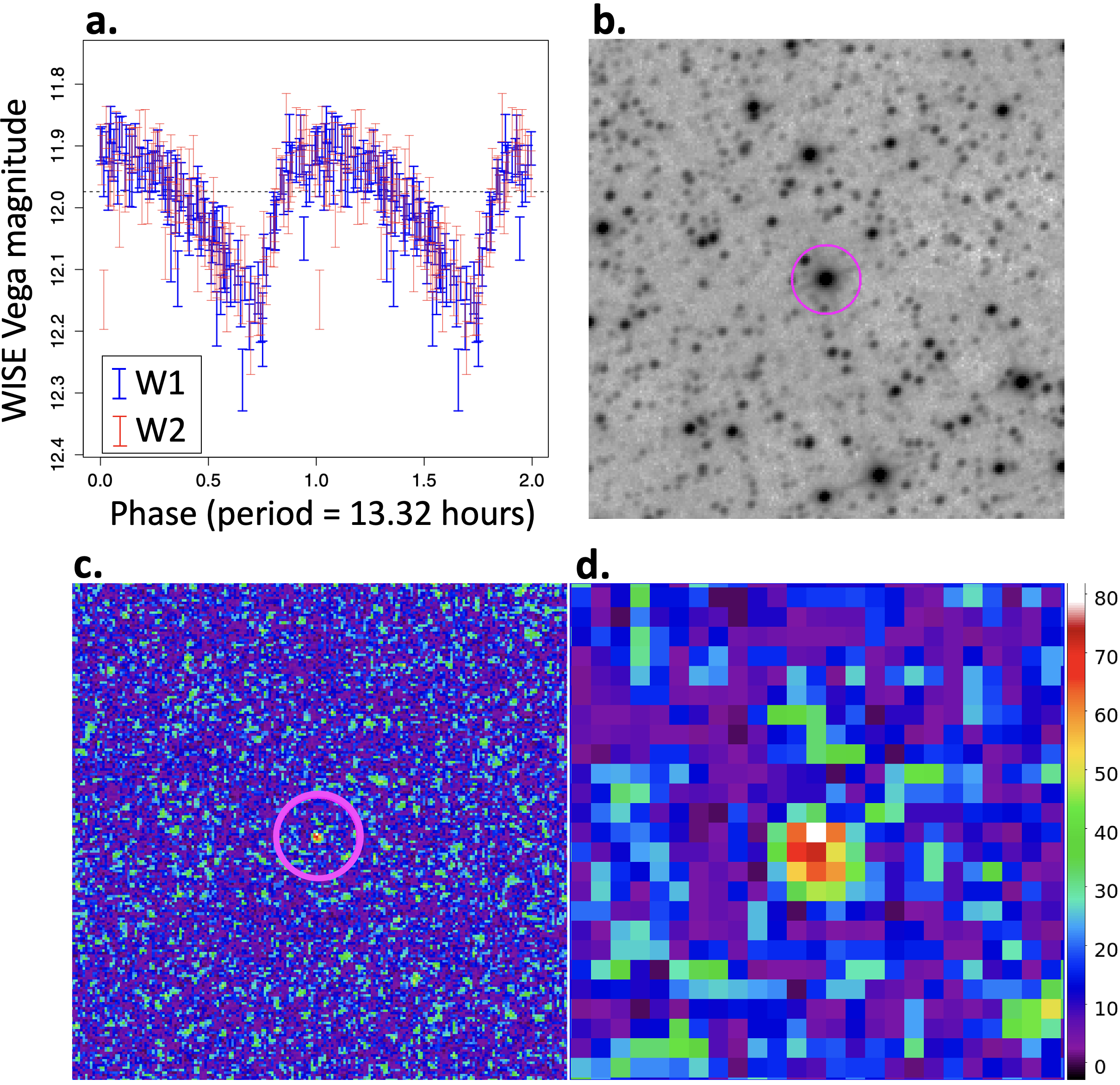}
  \caption{(a) Lightcurve of the RR Lyrae-type periodic variable star
           AVMen in the WISE W1 and W2 bands, phase-folded to a period of
           13.32 hours.
           (b) median image that collapses 174 WISE W1 single-exposure images
           centered on the location of AVMen. Image measures
           $\sim9.2\times 9.2$ arcmin$^2$. 
           (c) image of the CSDRL metric from collapsing all 174 exposures with           
           excess signal clearly detected.
           (d) Zoom-in on AVMen on the CSDRL metric image measuring
           $\sim1.15\times 1.15$ arcmin$^2$. The color-bar represents the range
           in CSDRL pixel values.}
  \label{fig:AVMen-w1}
\end{figure}

We tested the CSDRL method on the periodic variable star AV Men which is
detected with appreciable significance in the WISE W1 and W2 bands. This is
a RR Lyrae-type pulsator with a period of 13.32 hours. Its phase-folded
lightcurve is shown in Figure~\ref{fig:AVMen-w1}a. This was constructed from
WISE archival photometry by \cite{Masci2014}.
Following the recipe in Section~\ref{subsec:csdrl}, we applied a median
filter of length 3 timepoints to each pixel time series from $\sim$ 174 W1
time-ordered epochal images before computing the CSDRL image.
This smoothing maximized the overall signal (run-lengths) in pixels
centered on the location of AV Men relative to surrounding spurious
run-lengths resulting from noise fluctuations. The latter include the possible
systematics described above. The CSDRL image and a zoom-in on the variable
location are shown in Figures~\ref{fig:AVMen-w1}c and d. Pixels on the
peak location within a $3\times3$ pixel region have CSDRL values
of $\gtrsim 55$ compared to $\simeq 20$ for the background median value.
The longest runs contributing to the CSDRL signals are from the negative
pairwise slopes in the lightcurve where flux declines at a slower rate
and hence is better sampled by the WISE observing cadence.

Although not as significant as the results for the supernovae above,
this variable still shows up as the strongest spatially-correlated peak in
the CSDRL image. Based on the empirically-derived distribution of spurious
CSDRL cluster detections in this field, the probability of detecting the 
cluster of CSDRL values at the position of AVMen by chance is $<\sim 0.06\%$.
This upper bound is simply $100/N\left(\sum{CSDRL}\,<\,maxthres\right)$ where
$N\left(\sum{CSDRL}\,<\,maxthres\right)$ is the number of spurious clusters with
summed CSDRL pixel values less than the threshold {\it maxthres}, here set to 399,
-- the value measured for AV Men.
It is also worth noting that this field presents its own challenges due to a
relatively high level of source confusion. This confusion is seen to compound
systematics on/near the location of static (non-variable) sources that may be
blended with other sources. Despite this, the variable is still detected with
appreciable significance.

%------------------------------------------------------------------------------
\subsubsection{Future Improvements}

Our exploration of the CSDRL metric for different flavors of flux variability
has exposed some limitations that can be improved upon. The goal is to make this
metric as generic as possible for detecting a wide range of variability,
as limited by the available cadence, timespan of the input observations,
instrumental noise, and calibration errors. Some possible improvements and
extensions are as follows:

\begin{enumerate}
  \item{Automatically search for the optimal length of the median filter to
        use for mitigating noise prior to computing the CSDRL image, or rather,
        generate multiple CSDRL images for a range of input filter lengths.
        The input range to consider may be tied to prior knowledge of the cadence distribution
        (Section~\ref{subsec:histograms}), noise properties of the imaging data,
        but also characteristic timescales of the variability sought or
        admissible by a survey's design.}
  \item{Extend the CSDRL metric to combine the longest, 2nd longest,
        down to the $N$th-longest run in signed-slope in each
        pixel time series where $N$ is an input parameter. Another control
        parameter is to require that these runs differ in length by $\leq K$
        timepoints. This scheme will use more information from each time
        series and hence increase the statistical significance of the CSDRL
        values. This is particularly important when searching for sources
        with re-occurring flux variability (either periodic or aperiodic)
        since it combines multiple runs between successive peaks and troughs.
        For periodic variables, this is equivalent to phase-folding
        a lightcurve based on prior knowledge of its period, which maximizes
        the statistical significance for detecting flux variability
        exhibiting a specific characteristic shape.}
  \item{Generate images that encode the start and ending epochs in each
        pixel time series corresponding to the run-lengths used to compute
        the CSDRL metric image. This is useful for isolating the start and
        end of transient behavior within an input list of epochs.
        This information can then be used to generate lightcurves
        via forced photometry within timespans of interest.}
\end{enumerate} 
%---------------------End of Frank's subsection--------------------------------

\subsection{Tensor decomposition}
\label{subsec:tensors}

A third variability metric we explore is Tensor decomposition. Tensor decomposition can be used to decompose a set of images (i.e., a tensor) into different components. Similar to a classical Principal Component Analysis \citep[PCA, ][]{jackson2005user,hotelling1933analysis}, which composes a dataset into a set of multiple Eigenvectors, the Tensor Robust PCA \citep[TRPCA, e.g.,][]{LU2018} decomposes a tensor $\mathcal{X} \in \mathbb{R}^{n_1 \times n_2 \times n_3}$ into a \textit{constant low-rank} ($\mathcal{L}_0$) and a \textit{sparse} ($\mathcal{E}_0$) component such that $\mathcal{X} = \mathcal{L}_0 + \mathcal{E}_0$. This is achieved by minimizing the convex function
\begin{equation}\label{eq:tensor}
    \min_{\mathcal{L},\,\mathcal{E}} ||\mathcal{L}||_{*} + \lambda ||\mathcal{E}||_{1}, \, {\rm such\,that\,} \mathcal{X} = \mathcal{L}+\mathcal{E},
\end{equation}
where $||\mathcal{L}||_{*}$ denotes the nuclear norm (sum of the singular values of $\mathcal{L}$) and $||\mathcal{E}||_{1}$ denotes the $\ell_{1}$-norm (sum of the absolute values of all the entries in $\mathcal{E}$). Furthermore, $\lambda$ is related to a regularization hyper-parameter, which controls how much information is added to the sparse component (or vice versa to the low-rank component). As part of the training of the TRPCA, this parameter has to be tuned to a given dataset.

The TRPCA approach is ideal to recover a low-rank tensor corrupted by sparse errors. For example, the method is used to recover images with missing pixel values or to identify moving objects in security camera videos.
Motivated by the latter, here we test the performance of this technique to identify transient (in position or brightness) sources in a set of astronomical images. 

For this, we use an implementation of the above algorithm in the Python package \textsc{Tensorly} \citep[][]{tensorly}\footnote{\url{http://tensorly.org}}.
We apply TRPCA to a time series of $70$ \textit{WISE} 4.5$\mu$m (Band 2) cutouts of size $9.2\arcmin \times 9.2\arcmin$ (201$\times$201 pixels) around the variable star \textit{AV Men}, as
described in \ref{subsec:preprocess}. {\it AV Men} has a median brightness of $\sim$12\,mag and varies by about 30\% as shown in Figure \ref{fig:AVMen-w1}.
We use a learning rate of $1.2$, a maximal number of iterations of $20$, and a regularization for the low-rank and sparse component of $1.0$ and $0.17$, respectively. The latter serves similar to a weight and controls the amount of information added to the low-rank or sparse component. We found that decreasing the regularization for the sparse component results in more pixels being added to the sparse image component (hence also leading to more artifacts and spurious detections). In contrast, increasing the regularization of the sparse component results in only the most highly variable pixels being detected.

The regularization parameter for the sparse component therefore controls the completeness and reliability of the extracted transient candidate sample and needs to be fine-tuned for a given dataset. This can be done by optimizing a receiver operating characteristic (ROC) curve for different regularization parameters, based on the extraction of known transient sources, the insertion of mock transients into real data, or entirely or by entirely realistic end-to-end simulations of images.
%In our case, we choose parameters such as to minimize the inclusion of artifacts ({\it i.e.} non-variable sources).
%The latter are related to the ratio of detections in the low-rank and sparse components {\bf and need to be calibrated with simulations (e.g., insertion of transient objects to real data or entirely simulated images).}

\begin{figure}
    \centering
    \epsscale{0.5}
    \includegraphics[angle=0, width=6.0in]{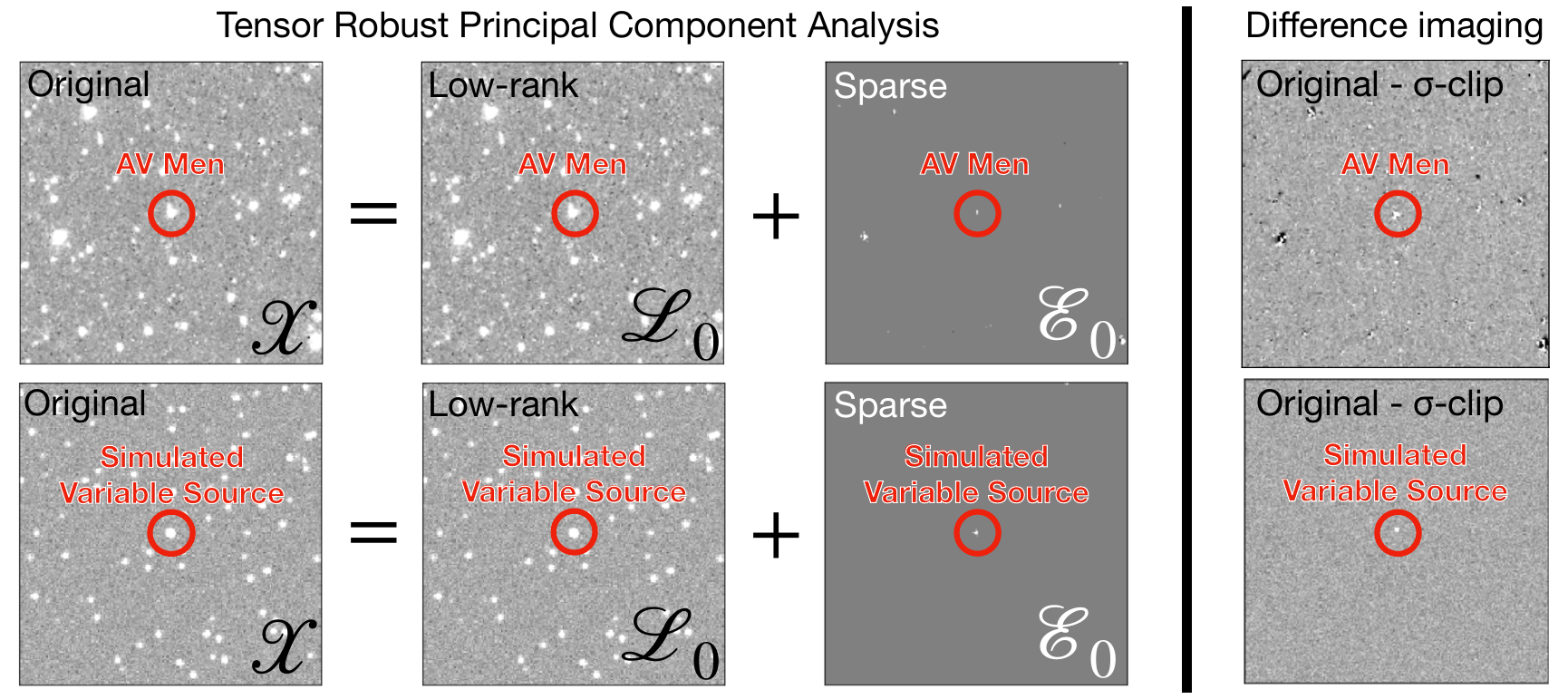}
    \caption{Example of the Tensor Robust Principal Component Analysis (TRPCA, panels on left) and ``classical'' difference imaging (right panels) on a 201$\times$201 pixel WISE image at 4.5$\mu$m. The top row shows the methods applied to our data (shown is one out of the $70$ frames). The variable star \textit{AV Men} is clearly identified in the sparse component, while the other non-variable sources appear in the low-rank component. TRPCA suppresses noise and extracts variable sources at a higher significance than difference imaging. However it also detects significant artifacts (likely due to changes in PSF or astrometric alignment between the frames) in the sparse component.
    The bottom row shows the same method applied on an idealized simulation (real noise, but static PSF and perfect astrometric alignment) with a mock variable source (similar properties as {\it AV Men} in the center. In this case the TRPCA method returns the location of the simulated variable source without other artifacts.
    %: here with X\% variability, the contrast is Y. %\red{Dave: can we indicate other variable sources on this frame? Maybe some of the ``artifacts'' are actually also variable}
    }
    \label{fig:tensor}
\end{figure}

The upper panels of Figure~\ref{fig:tensor} show the results of the decomposition for one of the $70$ frames centered on {\it AV Men} (left three panels, following Equation~\ref{eq:tensor}). The right panel shows the result of traditional difference imaging. With a very simple implementation of the TRPAC without much more testing on simulations, we are able to easily identify \textit{AV Men} in the sparse component image due to its variability between the different frames. Other objects, which are not variable over time, are part of the low-rank component.

Our test, however, also shows that {\it AV Men} is not uniquely identified in the sparse component. Instead, non-variable sources on the image are also detected by the TRPCA algorithm and appear on the sparse image. It is noticeable that these detections generally correspond to bright and extended sources. We think that they are likely being picked up by the algorithm due to variations in the PSF and/or differences in astrometric calibrations between the frames. 

We compared the results from the TRPCA with ``classical'' difference imaging, which is typically used to detect transients. The difference image was obtained by subtracting a $3\sigma-$clipped master frame from each of the original frames. Also with this method, {\it AV Men} is detected (upper right panel in Figure~\ref{fig:tensor}). We find that difference imaging is more susceptible to low-significance detections due to noise, while TRPCA results in a more bimodal detection distribution ({\it i.e.} a source is either detected as variable or not). We note that the other significant detections associated with the sparse component produced by the TRPCA also show up on the difference image, commonly as black-and-white pattern. These patterns are usually due to PSF variations or scatter in the astrometric alignment of the images. This confirms the above mentioned origin of spurious detections by the TRPCA.

%Generally, our relatively simple implementation of this approach easily identifies \textit{AV Men} due to its variability between the different frames.
%As expected, \textit{AV Men} appears in the sparse component, while the other objects, which are not transients, are part of the low-rank component.
%{\bf This test, however, also shows that some non-variable sources on the image are also detected by the TRPCA algorithm and appear on the sparse image. It is noticeable that these sources are generally bright and extended. We think that they are likely being picked up due to variations in the PSF and/or differences in the astrometric calibration between the frames.}

%in the difference image at the location of these other sources
%suggest that this is either due to PSF variation or scatter in the astrometric alignment of the images.}
%Clearly, the TRPCA is able to recover \textit{AV Men} at a higher significance. However, it also includes many artifacts that are not real variable sources. These are less significant (or negative pixels) in the difference image.
%\red{what are these artifacts??}

To further investigate the fidelity of the TRPCA method, we produced a set of $70$ simulated frames with realistic Poisson noises and background levels taken from the real images for each frame. To this, we add stationary ({\it i.e.} non-variable) sources at random positions as well as a variable source with similar variability and brightness as {\it AV Men} in the center of the image. For this simplistic simulation, we assumed point sources and a Gaussian PSF with ${\rm FWHM}=6.5\arcsec$, similar to the measured value for {\it W2}. Executing the same TRPCA method (using the same hyper-parameters), we are able to recover the location of the variable source without other artifacts (lower panels of Figure~\ref{fig:tensor}). This test shows that the method works assuming point sources, a constant PSF, and perfect astrometric alignment.
%we simulated a set of frames (how many?) with a fixed Gaussian PSF, including Poisson noise, and with an object whose brightness varied by X\%. Executing the same method, we are able to recover the variable source with a similar(?) sparse value as for the real data.

In conclusion, the TRPCA is a valid tool to identify variable objects in large astronomical images sets. Compared to classical difference imaging, the TRPCA method suppresses the background noise and therefore leads to more significant detection of candidate transient sources. However, the hyper-parameters need to be tuned carefully using real known transients or simulations (preferable end-to-end simulations) to optimize completeness and reliability, hence effectiveness, of the method. We found that changes in the PSF and astrometric alignment could lead to spurious detections.
One significant drawback of the TRPCA method is that it is relatively computationally intensive. It takes about $17\,{\rm seconds}$ to run the tensor composition on an $8$ core 2019 MacBook Pro in the above configuration.
%Furthermore, simulations are needed to calibrate the method (specifically the regularizations of the low-rank and sparse component detection) for optimal efficiency and performance.
The advantage of this method compared to, e.g., classical difference imaging is therefore not clear, yet.
%However, its advantage to classical methods such as difference imaging is not clear, yet. Another point to consider is that running the TRPCA on $70$ $201\,{\rm px} \times 201\,{\rm px}$ images {\bf in the above configuration} takes about $17\,{\rm seconds}$ on an $8$ core 2019 MacBook Pro. %Decreasing the maximal number of iterations decreases the CPU time linearly. We found that $10$ iterations lead to similar results in $7\,{\rm seconds}$. 

Our exploration of these metrics has revealed that if a source has a constant
baseline flux and then undergoes variability on top of
that baseline, the $\chi^2$ metric (Section~\ref{subsec:chi_squared}) results in a clean indicator of
variability that is relatively immune to the survey cadence. The CSDRL metric
(Section~\ref{subsec:csdrl}) however requires good temporal sampling of a lightcurve during
periods when the flux monotonically increases or decreases in excess of the
local noise. The shortest variability timescale detectable by the CSDRL metric
is therefore limited by the survey cadence.

\subsection{An Annotated Coadd Data Structure}

Based on our experience with these use cases we outline a potential data structure for an Annotated Coadd in Table \ref{datastructure}. The start time and the end time are the calendar time duration in MJD over which a sky coordinate was observed. The number of visits is the number of frames which were taken
at those coordinates. The sensitivity of a survey could vary from frame to frame because of exposure time, atmospheric conditions or sky background. In that scenario, we would generate a histogram of sensitivities per frame and provide the sensitivity percentiles.
The sampling cadence
column could either be the time-step bins derived from time differences between consecutive frames as described in Section 3.4 or the actual list
of MJD values of each frame - the latter could be prohibitively large to store for each pixel on the sky. These parameters are effectively survey description parameters and can be compiled on relatively large HEALPix pixels. 

On an individual source or pixel level, we would recommend storing the various variability metrics we have derived, since different metrics are sensitive to different kinds of variability as discussed above. We also suggest that the software derive a significance  value for each metric which can be thresholded  by the user to identify variability. 
Only sources or pixels above a specific threshold would be classified as robust variable sources, the equivalent of a signal to noise metric for the reliability of sources seen in a single frame. 

Finally, for each source identified to be variable, it could be desirable to generate and store the photometric light curve with forced photometry if a compact solution can be found in the data structure. However, since the number of candidate variable sources is a strong function of the significance threshold, the list of such sources is unlikely to be complete. Furthermore, since this is likely to result in a heterogeneous data product, strategies to optimize storage of this are deferred to future work.

Each of these parameters can be augmented as new data gets taken without having to revisit all the data from the past.

\begin{deluxetable}{cc}
\label{datastructure}
\tablecaption{Data Structure for an Annotated Coadd}
\tablehead{
\colhead{Name} & \colhead{Type of Data}\\
}
\startdata
Start time & HEALPix array\\
End time & HEALPix array\\
Number of visits & HEALPix array \\
Sensitivity percentiles per frame  & HEALPix arrays of magnitudes \\
Sampling cadence & HEALPix array, time-bin values\\
Variability Metric 1..n & 2D array in sky coordinates\\
Significance of variability metric 1..n & 2D array in sky coordinates \\
\enddata
\end{deluxetable}

%\subsubsection{Test of Tensor Decomp on moving sources}

%\subsection{Metrics in time windows} \label{subsec:rolling}
%Metrics can be computed in rolling windows over
%the time axis to accentuate transient features such
%as supernovae. 
%
%\subsubsection{Applying rolling window to SNe lightcurve}
%

\section{Conclusions} \label{sec:conclusions}

We have developed a set of tools and data products to simplify the identification of astrometrically and photometrically variable sources in time-domain sky surveys.
The tools track the number of epochs, the sensitivity and the temporal coverage of a particular data set on any region of sky. To ease access to this information for the entire sky, we have mapped this information onto a relatively coarse, HEALPix pixel grid. 
 We have then assessed three different quantitative criteria which work at the individual pixel level to encode the brightness time history using WISE, {\it Spitzer} and ZTF data. These are the reduced chi-squared metric, the consecutive-signed-difference run length metric and tensor principal component analysis, each of which condenses the brightness history over time of a pixel into a single metric. Together, the products are called an annotated coadd.
Of the three sample metrics, we find that:

\begin{enumerate}
    \item The reduced chi-squared metric, coupled with the number of samples in each sky pixel, is a powerful metric for a range of phenomena. Variable stars and slow movers (moving a fraction of a pixel between exposures) are readily distinguished with high significance. A drawback is that it is necessary to discard some data to remove artifacts, which also discards any fast movers (moving much more than a PSF-width between successive exposures). 
    \item The Consecutive Signed-Difference Run Length (CSDRL) metric works well
     when the variability timescale of interest is well sampled by the
     observations, i.e., are acquired at relatively high cadence. Pre-smoothing
     of the pixel lightcurves is also necessary to mitigate local noise
     fluctuations and improve overall signal-to-noise in the CSDRL metric. 
    \item Tensor PCA decomposition can separate moving objects from background when run on image differences, but so far we have not found advantages over simple thresholding on difference images. Moreover, the technique runs slowly as currently implemented. Since undiscovered transient phenomena will overwhelmingly be point sources, it is likely that working with point sources in difference images will be a superior technique to Tensor PCA.
\end{enumerate}

These metrics dramatically reduce by orders of magnitude, the volume of data that needs to be stored and accessed in a
time-critical way to select variable candidates for further follow-up. For example, knowing all past time-variable sources within the several square degree gravitational wave error regions, can 
help filter out unrelated sources that are detected. This will thereby
reduce significantly the list of
candidates whose lightcurves need to be reconstructed via forced photometry on the individual epochal images.
Future work will revolve around being able to account for
the difference in spatial resolution and slight differences in bandpasses between different data sets. We also present a data structure with the necessary information to ease the storage and search for temporally variable sources in future, large-volume datasets.

\begin{acknowledgements}
We thank the referee for a careful reading of the manuscript and for helpful suggestions.
\end{acknowledgements}

%% To help institutions obtain information on the effectiveness of their 
%% telescopes the AAS Journals has created a group of keywords for telescope 
%% facilities.
%
%% Following the acknowledgments section, use the following syntax and the
%% \facility{} or \facilities{} macros to list the keywords of facilities used 
%% in the research for the paper.  Each keyword is check against the master 
%% list during copy editing.  Individual instruments can be provided in 
%% parentheses, after the keyword, but they are not verified.

\vspace{5mm}
\facilities{Spitzer(IRAC), WISE, IRSA, ZTF}

%% Similar to \facility{}, there is the optional \software command to allow 
%% authors a place to specify which programs were used during the creation of 
%% the manuscript. Authors should list each code and include either a
%% citation or url to the code inside ()s when available.

\software{astropy \citep{2013A&A...558A..33A},
          numpy \citep{2020NumPy-Array},
          SciPy \citep{2020SciPy-NMeth},
          matplotlib \citep{2007Matplotlib},
          SWarp \citep[][]{bertin2010},
          Tensorly \citep[][]{tensorly}
          }

%% Appendix material should be preceded with a single \appendix command.
%% There should be a \section command for each appendix. Mark appendix
%% subsections with the same markup you use in the main body of the paper.

%% Each Appendix (indicated with \section) will be lettered A, B, C, etc.
%% The equation counter will reset when it encounters the \appendix
%% command and will number appendix equations (A1), (A2), etc. The
%% Figure and Table counter will not reset.

\bibliography{anncoadds}{}
\bibliographystyle{aasjournal}

%% Include this line if you are using the \added, \replaced, \deleted
%% commands to see a summary list of all changes at the end of the article.
%\listofchanges

\end{document}